\documentclass[12pt]{article}
\usepackage{amsmath,amsfonts,amssymb}
\usepackage[nosort]{cite}
\usepackage{hyperref}
\usepackage{graphicx}
\usepackage{color}
\usepackage{epsfig}
\usepackage{psfrag}	
\usepackage{tikz}
\usetikzlibrary{snakes}
\usepackage{slashed}
\usepackage{tensor}
\usepackage{ulem}
\usepackage[font=small,labelfont=bf,width=1\textwidth]{caption}
\usepackage{comment}
\usepackage[all]{xy}
\usepackage{multirow}
\usepackage{float}

\usepackage{color}

\usepackage{tocloft} % Layout of table of contents
\usepackage{ulem}
\usepackage{soul,xcolor}
\setstcolor{red}

\numberwithin{equation}{section}

\DeclareMathOperator{\Tr}{Tr}

\DeclareMathOperator{\sTr}{sTr}

\DeclareMathOperator{\diag}{diag}

\def\bC {\mathbb{C}}

\def\bP {\mathbb{P}}
\def\bR {\mathbb{R}}

\newcommand{\bea}{\begin{eqnarray}}
\newcommand{\eea}{\end{eqnarray}}
\newcommand{\beq}{\begin{equation}}
\newcommand{\eeq}{\end{equation}}
\newcommand{\bal}{\begin{equation}\begin{aligned}}
\newcommand{\eal}{\end{aligned} \end{equation}}
\newcommand{\half}{\frac{1}{2}}

\newcommand{\address}[1]{\vbox{\center\em#1}}
\renewcommand{\title}[1]{\vbox{\center\huge{#1}}\vspace{5mm}}

\newcommand{\cA}{{\mathcal A}}

\newcommand{\cD}{{\mathcal D}}

\newcommand{\cG}{{\mathcal G}}

\newcommand{\cL}{{\mathcal L}}

\newcommand{\cN}{{\mathcal N}}
\newcommand{\cP}{{\mathcal P}}

\newcommand{\su}{\mathfrak{s}\mathfrak{u}}
\newcommand{\sof}{\mathfrak{s}\mathfrak{o}}

\newcommand{\osp}{{\mathfrak{osp}}}
\newcommand{\uni}{{\mathfrak{u}}}

\newcommand{\Q}[3]{Q^{\dot{#1} #2 }_{#3}}
\newcommand{\barQ}[3]{Q^{\dot{#1} #2 }_{\bar{#3}}}

\begin{document}

\setstcolor{red}

\begin{titlepage}
\begin{center}

\vspace*{10mm}

\title{Conformal and non-conformal hyperloop deformations of the 1/2 BPS %\st{loop} 
circle}

\vspace{7mm}

\renewcommand{\thefootnote}{$\alph{footnote}$}

Nadav Drukker,%
\textsuperscript{1}
%\footnote{\href{mailto:nadav.drukker@gmail.com}
%{\tt nadav.drukker@gmail.com}}
Ziwen Kong,%
\textsuperscript{1}
%\footnote{\href{mailto:ziwen.kong@kcl.ac.uk}
%{\tt ziwen.kong@kcl.ac.uk}}
Malte Probst,%
\textsuperscript{1}
%\footnote{\href{mailto:malte.probst@kcl.ac.uk}
%{\tt malte.probst@kcl.ac.uk}}
\\
Marcia Tenser,%
\textsuperscript{2,3}
%\footnote{\href{mailto:marciatenser@gmail.com}
%{\tt marciatenser@gmail.com}}
and
Diego Trancanelli%
\textsuperscript{2,4}
%\footnote{\href{mailto:dtrancan@gmail.com}
%{\tt dtrancan@gmail.com}}

\vskip 2mm
\address{
\textsuperscript{1}%
Department of Mathematics, King's College London,
\\
The Strand, WC2R 2LS London, United-Kingdom}

\address{
\textsuperscript{2}%
Institute of Physics, University of S\~ao Paulo,
05314-970 S\~ao Paulo, Brazil
}

\address{
\textsuperscript{3}%
Dipartimento di Fisica, Universit\`a degli Studi di Milano-Bicocca, \\
Piazza della Scienza 3, 20126 Milano, Italy}

\address{
\textsuperscript{4}%
 Dipartimento di Scienze Fisiche, Informatiche e Matematiche, \\
Universit\`a di Modena e Reggio Emilia, via G. Campi 213/A, 41125 Modena, Italy \\ \& \\
INFN Sezione di Bologna, via Irnerio 46, 40126 Bologna, Italy}

\vskip .5cm
{\tt \{nadav.drukker, mltprbst, marciatenser, dtrancan\}@gmail.com\\ ziwen.kong@kcl.ac.uk}

\renewcommand{\thefootnote}{\arabic{footnote}}
\setcounter{footnote}{0}

\end{center}

\vspace{4mm}
\abstract{
\normalsize{
\noindent
We construct new large classes of BPS Wilson hyperloops in three-dimensional $\cN=4$ quiver 
Chern-Simons-matter theory on $S^3$. The main strategy is to start with the 1/2 BPS Wilson 
loop of this theory, choose any linear combination of the supercharges it preserves, and look for 
deformations built out of the matter fields that still preserve that supercharge. This is a powerful 
generalization of a recently developed approach based on deformations of 1/4 and 1/8 BPS bosonic 
loops, which itself was far more effective at discovering new operators than older methods relying on 
complicated ans\"atze. We discover many new moduli spaces of BPS hyperloops preserving varied 
numbers of supersymmetries and varied subsets of the symmetries of the 1/2 BPS operator. In 
particular, we find new bosonic operators preserving 2 or 3 supercharges as well as new families 
of loops that do not share supercharges with any bosonic loops, including subclasses of 
both 1/8 and 1/4 BPS loops that are conformal.}}
\vfill

\end{titlepage}

\tableofcontents

%%%%%%%%%%%%%

\section{Introduction and summary}

A distinguishing feature of three-dimensional supersymmetric conformal field theories are the vast moduli 
spaces of BPS line operators annihilated by some supercharges. For operators that are conformal, this was 
understood from an algebraic point of view in \cite{Agmon:2020pde}, but many examples of conformally 
invariant circular line operators, including continuous families of them, were found before, see for example 
\cite{Drukker:2008jm,Ouyang:2015qma,Cooke:2015ila, Ouyang:2015iza, Ouyang:2015bmy,Mauri:2017whf, 
Mauri:2018fsf, drukker2020bps, Drukker:2020dvr} and \cite{Drukker:2019bev} for a review. 

In the absence of an approach allowing for a full classification, we continue here to develop and employ 
constructive methods of identifying BPS Wilson loop operators called \emph{hyperloops}, finding a plethora 
of new observables, some of which are conformally invariant and some of which are not, greatly enlarging the 
known moduli spaces.

The theories we study are $\cN=4$ supersymmetric Chern-Simons-matter with either linear or circular quiver structure, characterized by the coupling of the gauge multiplet to hypermultiplets and twisted hypermultiplets \cite{Gaiotto:2008sd,Imamura:2008dt,Hosomichi:2008jd,Hama:2010av}. The 2-node circular quiver has $\cN=6$ supersymmetry and is the ABJ(M) theory \cite{Aharony:2008ug,Aharony:2008gk}, so most of what we say applies there as well. For concreteness, we consider theories on $S^3$ and focus on operators supported along a great circle.\footnote{Of course, it would be interesting to consider other contours, such as latitudes, or generic curves on an $S^2\subset S^3$, along the lines of what has been done in \cite{Drukker:2007dw,Drukker:2007yx,Drukker:2007qr} for ${\cal N}=4$ super Yang-Mills in four dimensions and in \cite{Cardinali:2012ru} for the ABJ(M) theory.}

In a recent paper \cite{Drukker:2020dvr}, some of us already studied Wilson loops in this same setting. 
Those hyperloops were written as deformations of bosonic Wilson loops that preserve 2 or 4 supercharges 
(so they are 1/8 or 1/4 BPS). Starting with particular block-diagonal combinations of bosonic connections 
$\cL_\text{bos}$ annihilated by a supercharge $Q$, it was found that one can deform them as follows
\beq
\label{bos-def}
\cL_\text{bos}\to\cL=\cL_\text{bos}-iQG+G^2,
\eeq
where $G$ is a matrix constructed out of bosonic fields in the hypermultiplets. The resulting operator is still supersymmetric, by construction, and is defined in terms of a superconnection containing the fermionic fields, which is something typical of supersymmetric Chern-Simons theories \cite{Drukker:2009hy}. Another peculiarity of three-dimensional theories is that the $Q$ variation of $\cL$ does not vanish {\it per se}, as it happens in the four-dimensional counterpart of these objects, but it is instead a total covariant derivative, so the entire Wilson loop, which is a gauge invariant object, is still annihilated by $Q$.

In the current work we apply a similar philosophy to \cite{Drukker:2020dvr}, but we employ as the starting point of the deformation the 
1/2 BPS Wilson loop found in \cite{Cooke:2015ila} (see also \cite{Ouyang:2015qma}), rather than a bosonic loop:
\beq
\label{ferm-def1}
\cL_{1/2}\to\cL=\cL_{1/2}+\textrm{deformation},
\eeq
with the details of the deformation given after \eqref{L} below. The 1/2 BPS loop is also a particular deformation 
of the bosonic loop as in \eqref{bos-def}, so our current construction includes all of those found previously. 

Moreover, unlike the construction in \cite{Drukker:2020dvr}, where a single choice of supercharge based on the 
original Wilson loops was employed, here we consider any supercharge annihilating the 1/2 BPS loop, so any 
linear combination of a basis of 8 supercharges. In particular, in cases when the supercharge $Q$ has an appropriate 
kernel, we find infinite-dimensional moduli spaces, since (roughly speaking) we can insert any of the operators in the 
kernel any number of times at any point along the loop.

This new procedure allows us to uncover new families of supersymmetric line operators. 
For example,  we have discovered: 
\begin{itemize}
\item 
Previously unrecognized bosonic loops preserving 2 and 3 supercharges, which are therefore 
1/8 and 3/16 BPS, in addition to the known ones preserving 2 or 4 supercharges, see Section~\ref{sec:bosonic}.

\item 
New 1/8 and 1/4 BPS loops that do not share supercharges with any 
known bosonic Wilson loops, so could not have been found by relying on \eqref{bos-def}. 
Of particular note is a subclass of these loops, which depends 
on one parameter (after fixing 4 supercharges), for which the variation of the superconnection 
under conformal transformations of the circle is a total derivative, see Section~\ref{sec:marcia1/4}.

This forms a new class of previously unrecognized line operators that are classically conformally 
invariant. Unlike the 1/2 BPS or 1/4 BPS bosonic loops, the one-dimensional conformal algebra is not generated by the supercharges 
that they preserve, but is an outer automorphism of it. As we cannot rely on supersymmetry to guarantee 
conformality, it would be extremely interesting to examine them at the quantum level and verify whether they  
are truly conformally invariant.
\end{itemize}

There are various natural directions that could be pursued starting from these results. The most obvious one is to try to compute the expectation value of these operators, using localization for example. This typically starts with determining to which cohomological class the various operators belong. In previous examples \cite{Drukker:2020dvr} based on \eqref{bos-def}, as well as in the original papers \cite{Drukker:2008jm,Cooke:2015ila}, it was found that the bosonic operators and their fermionic deformations are cohomologically equivalent. In this context we know however that this does not hold, as we find loops, such as the latitudes, that are known to have different expectation values from the 1/2 BPS circle \cite{Griguolo:2015swa,Bianchi:2016yzj,Bianchi:2016vvm,Bianchi:2018bke}. This of course makes these new classes of operators even more interesting.

The next natural question is about the holographic duals. While the holographic duals of 1/2 BPS loops in some ${\cal N}=4$ Chern-Simons-matter theories have been identified \cite{Ouyang:2015qma,Cooke:2015ila,Lietti:2017gtc}, the question  of what is dual to less supersymmetric (and/or higher representation) operators has not been 
addressed yet.%
\footnote{A first examination of a possible moduli space of 1/6 BPS loops in ABJ(M) theory was done in 
\cite{Correa:2019rdk,Correa:2021sky}.}

Finally, it would be interesting to study the moduli spaces of conformal loops as defect conformal manifolds and 
analyze the defect conformal field theory they define, along the lines of what has been done for the ABJ(M) 
theory in \cite{Bianchi:2020hsz} and see also \cite{Drukker:2022pxk}. For non-conformal loops it would be 
interesting to understand their renormalisation group flows \cite{Polchinski:2011im,Cuomo:2021rkm}.

This paper is organised as follows. In the next section we present the notation for the theories 
and the supersymmetry variations of the fields. In Section~\ref{sec:1/2} we present the simplest 1/2 BPS 
Wilson loop in these theories, which is the starting point of the deformations.
The bulk of the calculations is in Sections~\ref{sec:2-node} and~\ref{sec:longer}, focusing respectively 
on loops involving only two nodes of the quiver and those involving more, respectively. 
For the benefit of the casual reader we collect the main results and present a detailed analysis of 
special interesting examples in Section~\ref{sec:cases}. Some details are 
presented in the appendices.

%%%%%%%%%%%%%%

\section{${\cal N}=4$ Chern-Simons-matter theories on $S^3$}
\label{sec:notation}

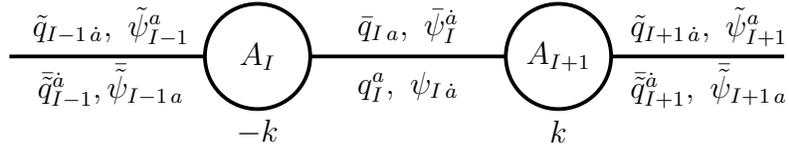
\begin{figure}[H]
\centering
\begin{tikzpicture}
\draw[line width=.5mm] (6,2) circle (7mm);
\draw[line width=.5mm] (10,2) circle (7mm);
\draw[line width=.5mm] (2.7,2)--(5.3,2);
\draw[line width=.5mm] (6.7,2)--(9.3,2);
\draw[line width=.5mm] (10.7,2)--(13,2);
\draw (6,2) node [] {$A_{I}$};
\draw (10,2) node [] {$A_{I+1}$};
\draw (6,1) node [] {$-k$};
\draw (10,1) node [] {$k$};
\draw (4.05,2.4) node [] {$\tilde{q}_{I-1\,\dot{a}}, \ \tilde{\psi}^a_{I-1}$};
\draw (4.05,1.6) node [] {$\bar{\tilde{q}}^{\dot{a}}_{I-1}, \bar{\tilde{\psi}}_{I-1\,a}$};
\draw (8,2.4) node [] {$\bar{q}_{I\,a}, \ \bar\psi^{\dot{a}}_{I}$};
\draw (8,1.6) node [] {$q^a_{I},\ \psi_{I\,\dot{a}}$};
\draw (12,2.4) node [] {$\tilde{q}_{I+1\,\dot{a}},\ \tilde{\psi}^a_{I+1}$};
\draw (12,1.6) node [] {$\bar{\tilde{q}}^{\dot{a}}_{I+1},\ \bar{\tilde{\psi}}_{I+1\,a}$};
\end{tikzpicture}
\caption{The quiver and field content of the ${\cal N}=4$ theory.}
\label{fig:N=4quiver}
\end{figure}

The theories we study are $\cN=4$ Chern-Simons-matter theories, which can be represented 
in terms of either circular or linear quiver diagrams 
 \cite{Gaiotto:2008sd,Imamura:2008dt,Hosomichi:2008jd,Hama:2010av}.
For the most part we focus on a node labeled by $I$ with gauge field $A_I$ and its adjacent 
node with $A_{I+1}$, but in Section~\ref{sec:longer} we also consider more nodes. 
The edges of the diagram represent hypermultiplets and twisted hypermultiplets. 
The hypermultiplet $(q_I^a,\psi_{I\dot a})$ couples to $A_I$ 
and $A_{I+1}$, while the twisted hypermultiplet $(\tilde q_{I-1\,\dot a},\tilde\psi_{I-1}^{a})$ couples 
to $A_I$ and $A_{I-1}$, and so on in an alternate fashion. 
The field content is summarized in the quiver diagram of Figure~\ref{fig:N=4quiver}, where 
the solid lines between nodes represent the matter fields.

The scalar fields in the hypermultiplet have indices $a,b=1,2$ and are doublets of the $SU(2)_L$ R-symmetry. 
The fermions with indices $\dot a,\dot b=\dot 1,\dot 2$ are charged instead under $SU(2)_R$. This is reversed 
in the twisted hypermultiplets. These indices are raised and lowered using the appropriate epsilon 
symbols: $v^a=\epsilon^{ab}v_ b$ and $v_a=\epsilon_{ab}v^b$ with $\epsilon^{12}=\epsilon_{21}=1$, 
and similarly for the dotted indices.

To write down the Wilson loops and the supersymmetry variations, it is useful to define 
moment maps and currents, following \cite{Cooke:2015ila,Drukker:2020dvr}
\bal
\mu_{I}{}^a_{\ b}&=q_I^a\bar q_{I\,b}-\frac{1}{2}\delta^a_bq_I^c\bar q_{I\,c}\,,
\qquad&
j_I^{a\dot b}&=q_I^a\bar\psi_I^{\dot b}-\epsilon^{ac}\epsilon^{\dot b\dot c}\psi_{I\,\dot c}\bar q_{I\,c}\,,
\\
\tilde\mu_{I}{}^{\dot a}_{\ \dot b}&=\bar{\tilde q}_{I-1}^{\,\dot a}\tilde q_{I-1\,\dot b}
-\frac{1}{2}\delta^{\dot a}_{\dot b}\bar{\tilde q}_{I-1}^{\,\dot c}\tilde q_{I-1\,\dot c}\,,
\qquad&
\tilde\jmath_I^{\,\dot ba}&=\bar{\tilde q}_{I-1}^{\,\dot b}\tilde \psi_{I-1}^{a}
-\epsilon^{\dot b\dot c}\epsilon^{ac}\bar{\tilde\psi}_{I-1\,c}\tilde q_{I-1\,\dot c}\,,
\\
\nu_{I}&=q_I^a\bar q_{I\,a}\,,
\qquad&
\tilde\nu_{I}&=\bar{\tilde q}_{I-1}^{\,\dot a}\tilde q_{I-1\,\dot a}\,.
\eal
These are bilinears of the matter fields and transform in the 
adjoint representation of $U(N_I)$.
Note that other bilinears of the same matter fields can transform in the adjoint of $U(N_{I\pm1})$. 
For example, $\nu_{I+1}=\bar q_{Ia}q_I^a$ is built out of the same fields as $\nu_I$, but it transforms in the adjoint of $U(N_{I+1})$ because of the reversed order. 

As stated in the Introduction, we define the theory on $S^3$ and the hyperloops we construct are 
supported along the equator of this sphere. The corresponding on-shell ${\cal N}=4$ supersymmetry 
transformations were derived in \cite{Drukker:2020dvr} by relying on the decomposition 
of $\cN=4$ to $\cN=2$ theories and the transformation rules of the latter in 
\cite{Hama:2010av,Asano:2012gt}. They are
\bal
\label{SUSY2}
\delta A_{\mu\,I}&=\frac{i}{k}\xi_{a\dot b}\gamma_\mu(j_I^{a\dot b}-\tilde\jmath_I^{\,\dot ba})\,,
\hskip-3cm\\
\delta q_I^a&=\xi^{a\dot b}\psi_{I\,\dot b}\,,
\qquad&
\delta\bar q_{I\,a}&=\xi_{a\dot b}\bar\psi_I^{\dot b}\,,\\
\delta \tilde q_{I-1\,\dot b}&=-\xi_{a\dot b}\tilde\psi_{I-1}^{a}\,,
\qquad&
\delta\bar{\tilde q}_{I-1}^{\,\dot b}&=-\xi^{a\dot b}\bar{\tilde\psi}_{I-1\,a}\,,
\hskip8cm\\
\delta\psi_{I\,\dot a}&=i\gamma^\mu\xi_{b\dot a}D_\mu q_I^b
+i\zeta_{b\dot a} q_I^b
-\frac{i}{k}\xi_{b\dot a}(\nu_I q_I^{b}-q_I^b\nu_{I+1})
+\frac{2i}{k}\xi_{b\dot c}\left(
\tilde\mu_{I}{}_{\ \dot a}^{\dot c}q_I^b
-q_I^b\tilde\mu_{I+1}{}_{\dot a}^{\ \dot c}\right),
\hskip-10cm
\\
\delta\bar\psi_{I}^{\dot a}&=i\gamma^\mu\xi^{b\dot a}D_\mu \bar q_{I\,b}
+i\zeta^{b\dot a}\bar q_{I\,b}
-\frac{i}{k}\xi^{b\dot a}(\bar q_{I\,b}\nu_I -\nu_{I+1}\bar q_{I\,b})
+\frac{2i}{k}\xi^{b\dot c}\left(
\bar q_{I\,b\,}\tilde\mu_{I}{}_{\ \dot c}^{\dot a}
-\tilde\mu_{I+1}{}_{\dot c}^{\ \dot a}\bar q_{I\,b}\right),
\hskip-11cm
\\
\delta\tilde\psi_{I-1}^{a}&=-i\gamma^\mu\xi^{a\dot b}D_\mu \tilde q_{I-1\,\dot b}
-i\zeta^{a\dot b}\tilde q_{I-1\,\dot b}
+\frac{i}{k}\xi^{a\dot b}(\tilde q_{I-1\,\dot b}\tilde\nu_{I} -\tilde\nu_{I-1}\tilde q_{I-1\,\dot b})
\hskip-10cm
\\&\quad
-\frac{2i}{k}\xi^{b\dot c}\left(
\tilde q_{I-1\,\dot c\,}\mu_{I}{}_{\ b}^{a}
-\mu_{I-1}{}_{b}^{\ a}\tilde q_{I-1\,\dot c}\right),
\hskip-1cm
\\
\delta\bar{\tilde\psi}_{I-1\,a}&=-i\gamma^\mu\xi_{a\dot b}D_\mu \bar{\tilde q}_{I-1}^{\,\dot b}
-i\zeta_{a\dot b}\bar{\tilde q}_{I-1}^{\,\dot b}+\frac{i}{k}\xi_{a\dot b}(\tilde\nu_I \bar{\tilde q}_{I-1}^{\,\dot b}-\bar{\tilde q}_{I-1}^{\,\dot b}\tilde\nu_{I-1})
\hskip-10cm
\\&\quad
-\frac{2i}{k}\xi_{b\dot c}\left(
\mu_{I}{}_{\ a}^{b}\bar{\tilde q}_{I-1}^{\,\dot c}
-\bar{\tilde q}_{I-1}^{\,\dot c}\mu_{I-1}{}_{a}^{\ b}\right),
%\hskip-11.5cm
\eal
where $\xi_{a\dot b}$ are the Killing spinors and 
$\zeta_{a\dot b}=\frac{1}{3}\gamma^\mu \nabla_\mu \xi_{a\dot b}$. The covariant derivative acts as, for instance, $D_\mu q^a_I=\partial_\mu q^a_I-iA_{\mu ,I}q^a_I+iq^a_I A_{\mu,I}$.

Specifically, each supersymmetry parameter $\xi^{a\dot b}$ is a linear combination of four 
(conformal) Killing spinors on $S^3$ denoted $\{\xi^l$, $\xi^{\bar l}$, $\xi^r$, $\xi^{\bar r}\}$, {\it i.e.}
\beq
\xi_{\alpha}^{a\dot b}=\xi_{\imath}^{a\dot b}\xi^\imath_\alpha + \xi_{\bar\imath}^{a\dot b}\xi^{\bar\imath}_\alpha\,,
\eeq
where $\imath = l, r$ and $\bar{\imath}=\bar l, \bar r$ label doublets of the 
$SO(2,1)$ conformal symmetry along the circle. All together they form a quartet of the $SO(4,1)$ symmetry of $S^3$. The index $\alpha=\pm$ is the spinor index.

The Killing spinors obey
\beq
\label{xi}
\nabla_\mu \xi^{l,\bar l} = \frac{i}{2}\gamma_\mu \xi^{l, \bar l}\,,
\qquad
\nabla_\mu \xi^{r, \bar r} = - \frac{i}{2}\gamma_\mu \xi^{r, \bar r}\,.
\eeq
Along the circle we take
$\gamma_\varphi=\sigma_3$ and the Killing spinors reduce to 
\cite{Assel:2015oxa}
\beq
\label{killingspinors}
\xi^l_\alpha=\begin{pmatrix}1\\0\end{pmatrix},
\qquad
\xi^{\bar l}_\alpha=\begin{pmatrix}0\\1\end{pmatrix},
\qquad
\xi^r_\alpha=\begin{pmatrix}e^{-i\varphi}\\0\end{pmatrix},
\qquad
\xi^{\bar r}_\alpha=\begin{pmatrix}0\\e^{i\varphi}\end{pmatrix},
\eeq
whence one finds $\zeta^{l,\bar l}_{a\dot b}=\frac{i}{2}\xi^{l,\bar l}_{a\dot b}$ and 
$\zeta^{r,\bar r}_{a\dot b}=-\frac{i}{2}\xi^{r,\bar r}_{a\dot b}$.

We work in Euclidean signature and take the gamma-matrices, $(\gamma^\mu)_\alpha^{\;\;\beta}$, to be given by the Pauli matrices. As usual, the spinor contractions are such that
\beq
\xi_1\xi_2\equiv \xi_1^\alpha \xi_{2,\alpha} =+\xi_ 2\xi_1\,,\qquad 
\xi_1 \gamma^\mu \xi_2\equiv 
\xi_1^\alpha (\gamma^\mu)_\alpha^{\;\;\beta} \xi_{2,\beta}
=-\xi_2 \gamma^\mu \xi_1\,.
\eeq
It follows that the Killing spinors in \eqref{killingspinors} satisfy 
$\xi^{\bar l}\xi^l=\xi^l \xi^{\bar l}=1$ 
and $\xi^{\bar l} \gamma^\mu \xi^{l}=-\xi^{l} \gamma^\mu \xi^{\bar l}=\delta^\mu_\varphi$, 
and similarly for the contractions involving $\xi^r$ and $\xi^{\bar r}$.

%%%%%%%%%%%%%%%%%%%%

\section{The 1/2 BPS Wilson loop and its symmetries}
\label{sec:1/2}

The starting point of the deformation \eqref{ferm-def1} considered in this paper is a particular 1/2 BPS loop of the theory. As shown originally for the ABJ(M) theory in \cite{Drukker:2009hy} and for $\cN=4$ theories in \cite{Cooke:2015ila} 
(see also \cite{Ouyang:2015qma}), such a Wilson loop must couple to at least 
two vector fields, as well as to the matter fields charged under them. We take the loop built around the 
$I$ and $I+1$ nodes of the particular form
\beq
\label{L1/2}
W_{1/2}=\sTr {\cal P} \exp i \oint\cL_{1/2}\,d\varphi\,,
\qquad
\cL_{1/2} = \begin{pmatrix}
\cA_{I}
& -i \bar\alpha \psi_{I \dot{1} -} \\
i \alpha \bar \psi^{\dot{1}}_{I+} 
&
\cA_{I+1}- \frac{1}{2}
\end{pmatrix},
\eeq
with
\beq
\cA_I=A_{\varphi, I} + \frac{i}{k}\big( {\nu_I} 
-\tilde \mu\indices{_I^{\dot{1}}_{\dot{1}}}
+\tilde \mu\indices{_I^{\dot{2}}_{\dot{2}}}\big)\,,\qquad
\cA_{I+1}=A_{\varphi, I+1} + \frac{i}{k} \big( {\nu_{I+1}}
-\tilde \mu\indices{_{I+1}_{\dot{1}}^{\dot{1}}}
+\tilde \mu\indices{_{I+1}_{\dot{2}}^{\dot{2}}}\big)\,.
\label{bosonicconnections}
\eeq
The constants $\alpha$ and $\bar\alpha$ (which are not complex conjugate to each other) satisfy $\alpha\bar\alpha=2i/k$ and the Wilson loop does not depend on their actual value, so we could fix them to be equal, but we leave them instead as a constant 
gauge parameter. We could allow for them to depend on $\varphi$ at the expense of a $U(1)$ gauge 
transformation at the bottom right entry: $\cA_{I+1}-\frac12\to\cA_{I+1}-\frac12-i\alpha^{-1}\partial_\varphi\alpha$. The origin of the shift $-1/2$ in the connection (and the resulting appearance of the supertrace if compared with the original definition in \cite{Cooke:2015ila} in terms of the trace) is explained in \cite{Drukker:2019bev}.

As we verify below, the eight supercharges preserved by this loop are 
\beq
\label{1/2susy}
Q^{\dot 2a +}_{\imath}
%Q^{\dot 2a\alpha}_{\imath}
\,,
\qquad 
Q^{\dot 1a -}_{\bar{\imath}}
%Q^{\dot 1a\alpha}_{\bar{\imath}}
\,.
\eeq
The spinor indices $\alpha=\pm$ are taken
upstairs, to contract with the downstairs indices of the Killing spinors in \eqref{killingspinors}. 
To relate to the notation in \eqref{SUSY2}, we can represent the supersymmetry transformation as
$\delta=-\xi^\imath_{a\dot b\alpha}Q^{\dot ba\alpha}_\imath-\xi^{\bar\imath}_{a\dot b\alpha}Q^{\dot ba\alpha}_{\bar\imath}$. 

Looking at the form of the Killing spinors along the circle \eqref{killingspinors}, one can write 
a general superposition of the preserved supercharges \eqref{1/2susy} as 
\beq
\label{Q}
Q = \eta^\imath_a Q^{\dot 2a +}_{\imath} 
+ \bar\eta^\imath_a (\sigma^1)_\imath^{\ \bar\imath} Q^{\dot1a -}_{\bar\imath}
= \eta^\imath_a\bar v_\imath Q^{\dot 2a +}
+ \bar\eta^\imath_a v_\imath Q^{\dot1a -}\,,
\eeq
with Grassmann-even parameters $\eta^\imath_a$, $\bar\eta^\imath_a$ (which, again, are not complex conjugate) and 
 auxiliary $SO(2,1)$ spinors
\beq
\label{vs}
v_\imath = \begin{pmatrix}e^{+i\varphi}\\1\end{pmatrix}_\imath\,,\qquad
\bar v_\imath = \begin{pmatrix}1\\e^{-i\varphi}\end{pmatrix}_\imath\,.
\eeq
The supersymmetry variations generated by a supercharge parameterised in such fashion can then be computed by reading off 
\beq
\label{SUSYpar}
\xi_{a{\dot 1}} = \begin{pmatrix} (\eta \bar v)_a \\ 0\end{pmatrix},
\qquad
\xi_{a{\dot 2}} = \begin{pmatrix} 0\\(\bar\eta v)_a \end{pmatrix}.
\eeq
In the right-most expression in \eqref{Q}, 
$Q^{\dot aa}$ acts in the same way as $Q^{\dot aa}_l$, that is without the extra phases $e^{\pm i \varphi}$, which have been absorbed in the definition of $v_\imath$ and $\bar v_\imath$. 
There are four $\eta^\imath_a$ and four $\bar\eta^\imath_a$ parameters, but the supercharges 
are identified up to rescalings, so the space of real supercharges is in fact $\bR\bP^7$.

As noted already in \cite{Cooke:2015ila}, there exists another Wilson loop with the same gauge fields and preserving the exact same symmetries, but coupling instead 
to other fields in the hypermultiplets. This other operator has the superconnection
\beq
\label{L1/2prime}
\cL_{1/2}' = \begin{pmatrix}
\cA_{I}
& -i \bar\alpha \psi_{I \dot{2} +} \\
i \alpha \bar \psi^{\dot{2}}_{I-} 
&
\cA_{I+1}+\frac{1}{2}
\end{pmatrix},
\eeq
with the opposite sign for the $\nu$'s compared to the ones appearing in \eqref{bosonicconnections}.
All the moduli spaces that we find include in them also this operator as a special point 
of enhanced supersymmetry. It is then just a matter of choice to do the analysis around \eqref{L1/2}, rather 
than around this one.

Before examining in detail the supersymmetries preserved by the loop defined in~\eqref{L1/2}, let us compute its bosonic symmetries. Our notation and further details on the algebra can 
be found in Appendix~\ref{app:SUSYtrans}. 
Firstly, notice that the superconnection \eqref{L1/2} contains only singlets of the $\su(2)_L$ R-symmetry, 
which is clearly preserved. The bosonic part of ${\cal{L}}_{1/2}$ is also annihilated by 
transverse rotations $T_\perp$, but it acts on the fermions by the Pauli matrix $\sigma_3$, see \eqref{Tperp}. Since 
spinor indices appear in ${\cal{L}}_{1/2}$ accompanied by opposite R-symmetry indices, 
we can cancel the action of $T_\perp$ by an appropriate multiple of the $\bar R_3$ generator of the unbroken 
$\uni(1)_R$ R-symmetry, and, indeed, the combination $L_\perp \equiv -i \left( T_\perp + i\bar R_3/2 \right)$
annihilates ${\cal{L}}_{1/2}$. As for the action of the conformal generators $J_0$ and $J_\pm$ on the $1/2$ 
BPS loop, using the expressions~\eqref{J0+-Bosons} and~\eqref{J0+-Fermions}
\bea
i J_0\, {\cal{L}}_{1/2} = \frac{d {\cal{L}}_{1/2}}{d \varphi} - \frac{\partial {\cal{L}}_{1/2}}{\partial \varphi} -[\sigma_3,{\cal{L}}_{1/2}]\,.
\eea
Since the ${\cal{L}}_{1/2}$ does not contain any explicit $\varphi$-dependence, we may bring this into the form%
\footnote{The precise definition of the covariant derivative ${\cal D}_\varphi^{{\cal L}_{1/2}}$ 
is in Appendix~\ref{app:SusyCondition}.} 
\bea
i J_0\, {\cal{L}}_{1/2} = {\cal{D}}^{{\cal{L}}_{1/2}}_\varphi \left( {\cal{L}}_{1/2} + \sigma_3 \right).
\eea 
Total covariant derivatives can be integrated away, so this guarantees invariance of the $1/2$ BPS loop under 
$J_0$. Similar arguments 
show that $J_\pm$ are preserved as well. Finally, note that while acting on ${\cal{L}}_{1/2}$ with 
$T_\perp$ (or equivalently $\bar R_3$) gives a non-zero result, it still takes the form of a covariant derivative
\bea
T_\perp {\cal{L}}_{1/2} \propto [\sigma_3, {\cal{L}}_{1/2}] = {\cal{D}}^{{\cal{L}}_{1/2}}_\varphi \sigma_3\,.
\eea
Consequently, $\bar R_3$ and $T_\perp$ are preserved separately. 

We now proceed to evaluate the action of the supercharge $Q$ in \eqref{Q} on the superconnection $\cL_{1/2}$ \eqref{L1/2} 
and to verify that it is equal to a total derivative. This also introduces a lot of the notation required 
in the rest of the paper.

First, to write the action of $Q$ on the hypermultiplet fields it is useful to define rotated scalar fields
\beq
\label{rotatedscalarfields}
r^1\equiv ( \eta \bar v)_a q^a \,,\quad
r^2\equiv ( \bar\eta v)_a q^a\,,\quad
\bar r_1\equiv \epsilon^{ab}(\bar\eta v)_a \bar q_b \,,\quad
\bar r_2\equiv -\epsilon^{ab}(\eta \bar v)_a \bar q_b\,,
\eeq
where $(\eta\bar v)_a=\eta^\imath_a \bar v_\imath$ and likewise for $(\bar\eta v)_a$. 
Now
\beq
\label{Qr}
Q r^1=-\Pi\psi_{\dot{2}+}\,,
\quad
Q r^2=\Pi\psi_{\dot{1}-}\,,
\quad
Q \bar r_1=\Pi\bar\psi^{\dot{2}}_-\,,
\quad
Q \bar r_2=-\Pi\bar\psi^{\dot{1}}_+\,,
\end{equation}
where the $\pm$ subscripts are spinor indices and
\bea
\label{Pi}
\Pi\equiv\epsilon^{ab}(\bar\eta v)_a( \eta\bar v)_b
\eea 
is a quantity that plays a central role in our analysis.

It is not too hard to show, using \eqref{SUSY2}, that the second variation of the rotated scalars is
\bal
Q^2 r^1&=
\Pi\left(i(\eta\bar v)_a \partial_\varphi q^a -\frac{1}{2}(\eta\sigma^3\bar v)_aq^a
+\cA_{I}r^1-\frac{2i}{k}\nu_{I}r^1-r^1\cA_{I+1}+\frac{2i}{k}r^1\nu_{I+1}\right),
\\
Q^2 r^2&=
\Pi\left(i(\bar\eta v)_a \partial_\varphi q^a -\frac{1}{2}(\bar\eta\sigma^3 v)_a q^a
+\cA_{I}r^2-r^2\cA_{I+1}\right).
\eal
Now, using
\bal
\label{sigma3}
2i\partial_\varphi(\eta\bar v)_a=(\eta\bar v)_a-(\eta\sigma^3\bar v)_a\,,\qquad
-2i\partial_\varphi(\bar\eta v)_a=(\bar\eta v)_a+(\bar\eta\sigma^3 v)_a\,,
\eal
and
\beq
\label{Pinu}
r^1\bar r_1+r^2\bar r_2=\Pi(q^1\bar q_1+q^2\bar q_2)=\Pi\nu\,,
\eeq
these second variations can be written as
\bal
\label{Q2r}
Q^2 r^1
&=
\Pi\left(i\partial_\varphi r^1-\frac{1}{2}r^1+\cA_{I}r^1-r^1\cA_{I+1}\right)
-\frac{2i}{k}(r^2\bar r_2r^1-r^1\bar r_2r^2)\,,
\\
Q^2 r^2&=
\Pi\left(i\partial_\varphi r^2 +\frac{1}{2}r^2
+\cA_{I}r^2-r^2\cA_{I+1}\right).
\eal
Likewise, the anti-chiral components have double variations given by
\bal
\label{Q2rbar}
Q^2 \bar{r}_1 &=\Pi \left(i \partial_{\varphi} \bar{r}_1 +\frac{1}{2} \bar{r}_1 
+\mathcal{A}_{I+1} \bar{r}_1 -\bar{r}_1 \mathcal{A}_{I} \right)
-\frac{2i}{k}(\bar r_2r^2\bar r_1-\bar r_1r^2\bar r_2)\,,
\\
Q^2 \bar{r}_2 &=\Pi \left(i \partial_{\varphi} \bar{r}_2 -\frac{1}{2} \bar{r}_2 
+\mathcal{A}_{I+1} \bar{r}_2 -\bar{r}_2 \mathcal{A}_{I} \right).
\eal

It is now straightforward to check that, when $\Pi\neq0$, the off-diagonal entries in $\cL_{1/2}$ are equal to $-i\Pi^{-1}QH$, with
\beq
\label{H}
H = \begin{pmatrix}
0 & \bar\alpha r^2 \\
 \alpha \bar r_2 & 0
\end{pmatrix}.
\eeq
One can combine this with the results above to find that the supersymmetry variation of the 1/2 BPS connection is
\beq
\label{QL1/2}
Q \cL_{1/2} =\mathcal{D}_{\varphi}^{\cL_{1/2}}H\,.
\eeq
The covariant derivative used here includes a commutator with the diagonal part of 
$\cL_{1/2}$ and an anticommutator with the off-diagonal part, as explained in detail in 
Appendix~\ref{app:SusyCondition}. 

For the purpose of this calculation it was not needed to evaluate the action of $Q^2$ on $r^1$, but only 
on $r^2$. The former is included here as it is of relevance for the rest of the paper. Also, if one wanted to 
repeat the calculation for the other 1/2 BPS loop in \eqref{L1/2prime}, one would need to replace 
$r^2$ and $\bar r_2$ in $H$ in \eqref{H} with $r^1$ and $\bar r_1$.

%%%%%%%%%%%%%%%%%%%%%

\section{Two-node hyperloops}
\label{sec:2-node}

Here we systematically study continuous deformations of the $\cL_{1/2}$ in \eqref{L1/2} 
 preserving the supercharge $Q$ defined in \eqref{Q}. Again, the strategy is not to find a superconnection which is strictly annihilated by $Q$, but that rather transforms as a total covariant derivative, precisely as ${\cal L}_{1/2}$ in \eqref{QL1/2} above. For the moment, we focus on the case in which the hyperloop couples to only two nodes of the quiver of the theory, but in the next section we generalize this to longer quivers.\footnote{The representation of the hyperloops in terms of quiver diagrams, which may include some or all of the nodes and edges of the original quiver defining the gauge theory, is explained in detail in \cite{drukker2020bps,Drukker:2020dvr}.}

Following \cite{Drukker:2020dvr}, we take a deformation of the form
\beq
\label{L}
\cL = \cL_{1/2} +F + B + C\,, 
\eeq
where $F$ is off-diagonal and Grassmann-odd, $B$ is a diagonal bilinear of the scalar fields and 
$C$ is annihilated by $Q$. This is the most general form consistent with the gauge group 
representations, the supermatrix structure and with all dimensions being equal to one. BPS non-conformal loops with higher dimension insertions are also possible, but are not considered here.

The condition $QC=0$ distinguishes two cases: when the supercharge annhilates some of the 
matter fields and when it does not. Nontrivial solutions include any BPS bosonic loop where the supersymmetry 
variation should be simply zero, rather than a total derivative. We exclude that case at the moment, 
because for a compact gauge group the coefficient of the gauge field in the Wilson loop is the identity 
(or more precisely $i$). As the gauge field already appears in the appropriate form in $\cL_{1/2}$, 
we should not allow for extra gauge field terms in $C$. An exception to this would arise if $B$ also 
has gauge fields, a possibility discussed in Appendix~\ref{app:modification}.

The other possibility is that $Q$ annihilates fields from the hypermultiplet.
Note that the action of $Q$ on the scalars in \eqref{Qr} is always proportional to the 
bilinear of the parameters $\eta^\imath_a$ and $\bar \eta^\imath_a$ that we called $\Pi$. When $\Pi$ is identically 
zero, we see that $Q$ has a nontrivial kernel (in this case $r^1\propto r^2$, so they do 
not form a basis of the scalar fields). One has therefore to distinguish the cases 
when $\Pi$ does not vanish (or has isolated zeros) and the case when $\Pi=0$, studied later in 
Section~\ref{sec:2-node-0}.

%%%%%%%%%%%%%%%%

\subsection{Deformations with $\Pi\neq0$}
\label{sec:2-node-not0}

Starting from the ansatz \eqref{L}, we want to determine the most general $F$, $B$ and $C$ 
giving BPS loops, under the assumption that $\Pi\neq 0$.

The simplest term to address is $C$. The only solutions to $QC=0$ which is at most bilinear in the fields and 
excluding the gauge field is $C=\diag(c_I,c_{I+1})$, a numerical matrix not containing the fields. 
Note that we set the radius $R$ of 
$S^3$ to 1, otherwise this should scale with $1/R$ on dimensional grounds. The term proportional 
to the identity is completely trivial, so we remove it and take $C=\diag(0,c)$.

Moving on to $F$, in order for $QF$ to involve a derivative in the $\varphi$ direction, 
$F$ is restricted to have the fermions in \eqref{Qr}. Therefore, if $\Pi\neq0$, one can take
\beq
F=-iQG\,,\qquad
G = \begin{pmatrix}
0 & \bar b_a r^a \\
b^a \bar r_a & 0\end{pmatrix}.
\label{FiQG}
\eeq
Here the $b^a$, $\bar b_a$ parameters may be functions of $\varphi$.

In terms of $G$, we can combine \eqref{Q2r} and \eqref{Q2rbar} into
\begin{equation}
\label{Q2G}
{-i}Q^2 G =\partial_{\varphi} (\Pi G) -i[\cL_{1/2}^B, \Pi G] +i[H^2,G] -\Pi\hat\cG\,,
\end{equation}
with the remainder
\beq
\Pi\hat\cG= 
\begin{pmatrix}
0 & \partial_\varphi(\Pi\bar b_a) r^a -i\Pi \bar{b}_1 r^1\\
\partial_\varphi(\Pi b^a) \bar r_a+i\Pi b^1 \bar r_1 & 0\end{pmatrix}.
\eeq
To evaluate the supersymmetry variation, it is sometimes useful to split the connection into the diagonal 
(bosonic) and off-diagonal (fermionic) part: $\cL=\cL^B+\cL^F$, and likewise for $\cL_{1/2}$.
One can then write
\bal
\label{QcL}
Q\cL&=Q\cL_{1/2}-iQ^2G+QB
\\&=
\cD^{\cL_{1/2}}H+\partial_{\varphi} (\Pi G) -i[\cL_{1/2}^B, \Pi G] +i[H^2,G]+QB-\Pi\hat\cG
\\&=
\cD^{\cL_{1/2}}(H+\Pi G)-i\{\cL_{1/2}^F, \Pi G\} +i[H^2,G]+QB-\Pi\hat\cG
\\&=
\cD^{\cL}(H+\Pi G)+i[B,H+\Pi G]+i[C,H+\Pi G]+i[H^2,G]-\Pi\hat\cG
\\&\quad-\{QG,H+\Pi G\}-\{QH,G\}+QB\,.
\eal
The terms on the last line are all diagonal and vanish by setting $B=\{G, H\} +\Pi G^2$. With this form for $B$, also 
the second and fourth terms on the previous line (which are cubic in the scalar fields) vanish.
Another way to write these equations is in terms of the variations of the extra terms in 
$\cL$ in \eqref{L}
\begin{equation}
\begin{aligned}
\label{BF-var}
{Q} B &=i\{F,H\} +\{\mathcal{L}_{1/2}^F +F,\Delta H\}\,,\\
{Q} F &=\partial_{\varphi} \Delta H -i[\mathcal{L}_{1/2}^B, \Delta H] -i[B+C, H+\Delta H]\,.
\end{aligned}
\end{equation}
We see that this is indeed satisfied with $\Delta H=\Pi G$.

The deformed connection can then be written as
\begin{equation}
\label{deformed-L}
\cL= \cL_{1/2} -iQ G +\{G, H\} +\Pi G^2 +C\,,
\end{equation}
and it is a total derivative if we further impose that the remainders in the last equality of \eqref{QcL} cancel
\beq
i[C,H+\Pi G]-\Pi\hat\cG=0\,.
\eeq
These are four differential equations for $b^a$ and $\bar b_a$
\begin{equation}
\label{4diffeq}
\begin{aligned}
\partial_{\varphi} (\Pi b^1) -i(c-1) \Pi b^1&=0\,, \\
\partial_{\varphi} (\Pi b^2)-ic (\alpha +\Pi b^2)&=0\,,\\
\partial_{\varphi} (\Pi \bar{b}_1) +i (c-1) \Pi \bar{b}_1&=0\,, \\
\partial_{\varphi} (\Pi \bar{b}_2) +i c (\bar{\alpha} +\Pi \bar{b}_2)&=0\,.
\end{aligned}
\end{equation}
Taking $\hat c(\varphi)$ to be the primitive of $c$, the general solution can be written as
\beq
\label{beta}
\Pi b^1 =e^{-i\varphi+i\hat c}\beta^{1}\,, 
\quad 
\Pi b^2 = e^{i\hat c}\beta^{2}-\alpha\,,
\quad
\Pi\bar b_1 = e^{i\varphi-i\hat c}\bar\beta_1 \,, 
\quad 
\Pi\bar b_2 = e^{-i\hat c}\bar\beta_2-\bar\alpha\,,
\eeq
with constant $\beta^1$, $\beta^2$, $\bar\beta_1$, $\bar\beta_2$.

There is a lot of freedom in choosing $c$. It can in principle be an arbitrary function of $\varphi$, 
but this is a gauge symmetry, which is absorbed in $\cA_{I+1}$. We can always fix to the
same gauge as in \eqref{L1/2} by setting $c=0$. Note that in generic gauges, when $\hat c$ is not 
periodic, the parameters $b$ and $\bar b$ are also not periodic (as it was in the original 
paper \cite{Drukker:2009hy}).

In the gauge $c=0$, the deformed connection \eqref{deformed-L} is
\begin{equation}
\label{L-explicit}
\cL=\begin{pmatrix}
A_{\varphi,I} + M_a^{\ b} r^a \bar{r}_b 
-\frac{i}{k} (\tilde{\mu}_{I \ \dot{1}}^{\ \dot{1}} - \tilde{\mu}_{I \ \dot{2}}^{\ \dot{2}}) 
& -i \bar{\beta}_2 \psi_{I \dot{1}-} +ie^{i\varphi}\bar{\beta}_1 \psi_{I \dot{2}+}\\
i\beta^{2} \bar{\psi}_{I+}^{\dot{1}} -ie^{-i\varphi}\beta^1 \bar{\psi}_{I-}^{\dot{2}}
& A_{\varphi, I+1} +M_{a}^{\ b} \bar{r}_b r^a -\frac{i}{k} (\tilde{\mu}_{I+1\, \dot{1}}{}^{\dot{1}} - \tilde{\mu}_{I+1\, \dot{2}}{}^{\dot{2}}) - \frac{1}{2}
\end{pmatrix},
\end{equation}
where
\begin{equation}
\label{beta-M}
M=\Pi^{-1}\begin{pmatrix}
\bar{\beta}_1 \beta^{1} +\frac{i}{k} & e^{i\varphi}\bar{\beta}_1 \beta^{2}\\
e^{-i\varphi}\bar{\beta}_2 \beta^{1} & \bar{\beta}_2 \beta^{2} -\frac{i}{k}
\end{pmatrix}.
\end{equation}
After fixing a supercharge $Q$, the possible space of hyperloops it generates can be represented 
by the matrix $M$ in \eqref{beta-M}. It is given by 4 complex parameters $\beta^a$ and $\bar\beta_a$, modded out 
by a $C^* $ action, which is the conifold. This is the same type of moduli space found in 
\cite{Drukker:2019bev, Drukker:2020dvr}.

Note that the effect of the shift of $\beta^2$ and $\bar\beta_2$ by $\alpha$ and $\bar\alpha$ in \eqref{beta} 
means that the ``origin of $\beta$ space'', which is the tip of the conifold, is a bosonic loop. We can thus view all 
the hyperloops that we find here as deformations around some bosonic loop by some supercharge that it preserves. 
This is similar to the structure in \cite{Drukker:2020dvr}, but here we have far more general bosonic 
loops (see Section~\ref{sec:bosonic}) and choose any of the supercharges that they preserve.

Specific examples of hyperloops of this type are presented in Section~\ref{sec:cases2nodes}. Their 
symmetry algebras are also studied there, as well as a closer inspection of the connection between them 
and the hyperloops of \cite{Drukker:2020dvr}.

%%%%%%%%%%%%

\subsubsection{The condition $\Pi\neq 0$ from an algebraic point of view}
\label{sec:alg1}

The conditions on $\Pi$ being zero or not can be interpreted from an algebraic point of view. To do that, let us start by looking at the square of the  supercharge \eqref{Q}, which using \eqref{anticomms} reads
\begin{align}
\label{QQ'}
Q^2 = - \Pi_-  J_- - \Pi_0 J_0 + \Pi_+ J_+ - \lambda L_\perp - \half \lambda_{ab} R^{ab}\,,
\end{align}
with $\Pi_{\pm}$ and $\Pi_0$ the Fourier coefficients of $\Pi$, defined through
\beq
\label{Piab}
\Pi \equiv \Pi_- e^{-i\varphi} + \Pi_0 + \Pi_+ e^{+i\varphi} \,,
\eeq
and
\beq
\label{lambdaab}
\lambda_{ab} \equiv \epsilon_{\imath\jmath} \bar\eta^\imath_a \eta^\jmath_b\,, 
\qquad 
\lambda \equiv \epsilon^{ab} \lambda_{ab}\,.
\eeq
As mentioned in Section~\ref{sec:1/2}, $J_0$ and $J_\pm$ are the generators of the conformal group along the circle, $R^{ab}$ are $\su(2)_L$ generators, 
and $L_\perp$ is a combination of rotation orthogonal to the circle and the unbroken $\uni(1)_R$ 
(see Appendix~\ref{app:SUSYtrans} for further details). 

As manifest from \eqref{QQ'}, $Q^2$ generically generates 
$\mathfrak{sl}(2,\mathbb{R}) \oplus \su(2)_L \oplus \uni(1)_{L_\perp}$, which is the algebra preserved 
by the 1/2 BPS Wilson loop. 
When $\Pi \neq 0$ the conformal generators are part of this preserved algebra (at least in part). It is 
now possible to consider subcases of the condition $\Pi\neq 0$ in which one progressively decouples 
some of the generators on the right hand side of \eqref{QQ'}. This imposes conditions on the parameters $\eta$ and $\bar\eta$, which we derive below and which are going to be  
useful in Section~\ref{sec:cases}, where we construct specific examples.

We start by considering cases in which the $\su(2)_L$ is ``turned off''. In order for the contribution 
of $R^{ab}$ to $Q^2$ to vanish, one must 
require that $\lambda_{ab}$ in (\ref{lambdaab}) be antisymmetric. This implies that
\bea
\lambda_{11} = \epsilon_{\imath\jmath}\bar\eta^\imath_1 \eta^\jmath_1=0\,,
\eea
which allows to deduce $\bar\eta^\imath_1 \propto \eta^\imath_1$, and similarly for $\lambda_{22}$ and $\bar\eta^\imath_2, \eta^\imath_2$. We may then factorize these parameter in terms of some other quantities carrying a single index, as follows (bars do not indicate complex conjugation, as usual)
\bal
\bar\eta^\imath_1 &= \bar w_1 s^\imath\,, 
&\qquad
\eta^\imath_1 &= w_1 s^\imath\,, 
\\
\bar\eta^\imath_2 &= \bar w_2 t^\imath\,,
&\qquad
\eta^\imath_2 &= w_2 t^\imath.
\eal
It remains to impose 
\bal
\lambda_{12} + \lambda_{21} = (\epsilon_{\imath\jmath} s^\imath t^\jmath) (\epsilon^{ab} \bar w_a w_b)=0\,,
\eal
which can be achieved by setting either $s^\imath \propto t^\imath$ or $\bar w_a \propto w_a$.
As a consequence, the remaining parameters that determine $Q^2$ are given by 
\beq
\lambda = (\epsilon_{\imath\jmath} s^\imath t^\jmath) (\bar w_1 w_2 + \bar w_2 w_1)\,,
\qquad
\Pi = \left( s^l e^{+i \varphi/2} + s^r e^{-i \varphi/2} \right)^2\epsilon^{ab} \bar w_a w_b\,.
\eeq
In order to avoid that $\Pi = 0$, we must ensure $\epsilon^{ab} \bar w_a w_b \neq 0$, which implies $\epsilon_{\imath\jmath} s^\imath t^\jmath =0$.
In particular, the contribution of $L_\perp$ vanishes automatically. In other words, 
$Q^2 \in \mathfrak{so}(2,1)$. More restrictive cases can be easily constructed by considering special choices 
of $s^l$, $s^r$. In particular, setting $s^r = 0$ gives $Q^2 \propto J_+$ and similarly $s^l=0$ gives $Q^2 \propto J_-$.

Next, one could maintain the $\su(2)_L$ and set instead individual Fourier 
coefficients of $\Pi$ to zero, looking, for example, to the case 
$Q^2 \in \mathfrak{u}(1)_{J_0} \oplus \su(2)_L \oplus \mathfrak{u}(1)_{L_\perp}$. 
The contributions of $J_\pm$ to $Q^2$ vanish if and only if
\begin{align}
\label{lastcase}
\epsilon^{ab} \eta^l_a \bar\eta^l_b=0\,,
\qquad
\epsilon^{ab} \eta^r_a \bar\eta^r_b=0\,,
\end{align}
namely if the $\eta$'s are linearly dependent
\bal
\label{etatw-etatz}
\eta^l_a &= t^l w_a\,, 
\qquad&
\eta^r_a &= t^r z_a\,, \\ 
\bar\eta^l_a &= \bar t^l w_a\,,
\qquad&
\bar\eta^r_a &= \bar t^r z_a\,.
\eal
Without loss of generality, one can take $z,w$ to be normalized, finding the corresponding parameters
\bal
\label{PiLambdaNonconformal}
\Pi &= ( \epsilon^{ab} z_a w_b) \, (\epsilon_{\imath\jmath} \bar t^\imath t^\jmath) \,, \\
\lambda_{ab} &= (\epsilon_{\imath\jmath} \bar t^\imath t^\jmath ) z_{(a} w_{b)} + \half \bar t^\imath \begin{pmatrix} 0 & 1 \\ 1 & 0 \end{pmatrix}_{\imath\jmath} 
 t^\jmath (\epsilon^{cd} z_c w_d) \epsilon_{ab} \,.
\eal

One could go on and, for example, turn off $\Pi_-$ and $\Pi_0$ by imposing
\bal
0 = \epsilon^{ab} \bar\eta^{r}_a \eta^r_b\,, 
\qquad 
0 = \epsilon^{ab} \bar\eta^{r}_a \eta^l_b + \epsilon^{ab} \bar\eta^{l}_a \eta^r_b\,,
\eal
which is achieved by taking
\bal
\label{etasw}
\eta^r_a = s w_a,
\qquad
\bar\eta^r_a = \bar s w_a\,, 
\eal
and yields
\bal
\label{LambdaNilpotent}
\lambda_{ab} %&= \epsilon_{\imath\jmath} \bar\eta^\imath_a\eta^\jmath_b \\
%&= \bar\eta^l_a \eta^r_b - \bar\eta^r_a \eta^l_b \\
%&= \left( \bar s \eta^l_a - t w_a \right) w_b - \bar s w_a \eta^l_b \\
&= \bar s \left( \eta^l_a w_b - w_a \eta^l_b \right) - t w_a w_b\,.
\eal
%and in particular $\lambda_{(ab)} = -t w_a w_b$.

The specific cases considered above do not form an exhaustive classification of supercharges with $\Pi \neq 0$, but have been selected because they are of interest in the study of some loops, like the bosonic loops in Section~\ref{sec:bosonic}. Supercharges whose squares are a linear combination of both $\su(2)_L$ and conformal generators can nonetheless be easily constructed.

%%%%%%%%%%%%%%%%%%%%%%%

\subsection{Deformations with $\Pi=0$}
\label{sec:2-node-0}

The analysis above gives Wilson loops rather similar to those already studied in \cite{Drukker:2020dvr} 
(though far more general). As seen, it requires that the function $\Pi$ be non-zero. Now we turn to look 
at the interesting case when
\begin{equation}
\label{Pi=0}
\Pi=(\bar{\eta} v)_1 (\eta \bar{v})_2 -(\bar{\eta} v)_2 (\eta \bar{v})_1 =0\,,
\end{equation}
and define
\begin{equation}
\label{xipar}
\xi =\frac{(\eta \bar{v})_1}{(\bar{\eta} v)_1} =\frac{(\eta \bar{v})_2}{(\bar{\eta} v)_2}\,,
\end{equation}
thus $\xi(\varphi)\in \mathbb{C}\cup \{\infty\}$.

This case is subtle because the supercharge $Q$ in \eqref{Q} annihilates the rotated scalars \eqref{Qr} 
and, furthermore, the pairs of rotated fields are not linearly independent
\begin{equation}
r^1=\xi r^2\,,\qquad \bar{r}_2=-\xi\bar{r}_1\,.
\end{equation}

For convenience we define (assuming $(\bar{\eta}v)_1 \ne 0$)
\beq
\label{rparallel}
r^\parallel=r^2\,,\qquad \bar r_\parallel=-\bar r_1\,,
\eeq
and an orthogonal pair which are not annihilated by $Q$
\begin{equation}
r^{\perp} = (\bar{\eta} v)_2 q^1 -(\bar{\eta} v)_1 q^2\,,\qquad
\bar{r}_{\perp} =(\bar{\eta} v)_1 \bar{q}_1 + (\bar{\eta} v)_2 \bar{q}_2\,.
\end{equation}
We then find that
\begin{equation}
\begin{aligned}
\label{Pi0Qr}
%{Q} r^{\parallel}={Q} \bar{r}_{\parallel} =& 0,\quad 
{Q} r^{\perp} &= \Lambda (\xi \psi_{\dot{1}-} +\psi_{\dot{2}+})\,,\qquad
{Q} \bar{r}_{\perp} =-\Lambda (\bar{\psi}_+^{\dot{1}} -\xi \bar{\psi}_-^{\dot{2}})\,,\\
{Q}^2 r^{\perp} &= -\Lambda \left((i\partial_{\varphi} \xi -\xi)r^{\parallel} -\frac{2i}{k} \xi (\nu_{I} r^{\parallel} - r^{\parallel} \nu_{I+1})\right),\\
{Q}^2 \bar{r}_{\perp} &= \Lambda \left((i\partial_{\varphi} \xi -\xi)\bar{r}_{\parallel} +\frac{2i}{k} \xi (\nu_{I+1} \bar{r}_{\parallel} -\bar{r}_{\parallel} \nu_I)\right),
\end{aligned}
\end{equation}
where 
\bea
\Lambda\equiv (\bar{\eta}v)_1^2 +(\bar{\eta}v)_2^2\,,
\eea
and similarly to \eqref{Pinu}
\begin{equation}
\label{Pi0nu}
r^{\parallel} \bar{r}_{\perp} +r^{\perp} \bar{r}_{\parallel} =\Lambda \nu_I\,,
\qquad 
\bar{r}_{\perp} r^{\parallel} +\bar{r}_{\parallel} r^{\perp}=\Lambda \nu_{I+1}\,.
\end{equation}

We can apply now the same formalism as in the $\Pi\neq0$ case and take
\begin{equation}
\label{Q=0L}
\mathcal{L} =\mathcal{L}_{1/2} -i{Q} G +\{G, H\} +C\,,\qquad QC=0\,.
\end{equation}
$H$ is the same as above, see \eqref{H}, which in the new notations becomes
\beq
\label{Pi0H}
H = \begin{pmatrix}
0 & \bar\alpha r^\parallel \\
 \alpha \xi\bar r_\parallel & 0
\end{pmatrix}.
\eeq
In $G$ we include only $r^\perp$ and $\bar r_\perp$ and $C$ 
may contain scalar bilinears as well as the numerical factors discussed before
\begin{equation}
\label{Pi=0G}
G=\begin{pmatrix}
0 & \bar{\beta}_\perp r^{\perp}\\
\beta^\perp \bar{r}_{\perp} & 0
\end{pmatrix},
\qquad
C=\begin{pmatrix}
\bar{\beta}_\parallel r^{\parallel}\bar r_{\parallel}&0\\
0&\beta^\parallel \bar{r}_{\parallel}r^\parallel+c
\end{pmatrix}.
\end{equation}
$QG$ gives a single linear combination of the fermions $\xi\psi_{\dot{1}-} +\psi_{\dot{2}+}$. In 
Appendix~\ref{app:modification} we explore the possibility of adding another combinations of the 
fermions, but find that this can only be done in the case of $\xi=0$, presented in 
Section~\ref{sec:2-node-xi} below.

Going back to the deformation \eqref{Pi=0}, one can get $Q\cL=\cD_\varphi^\cL H$, provided that
\begin{equation}
\label{pi=0susycondition}
Q^2 G= [G,H^2] +[C, H]\,.
\end{equation}
Unlike the $\Pi\neq0$ case, here $H$ remains the same regardless of the deformation.

Besides, one can check that the cubic terms inside $Q^2 G$ cancel 
$[G,H^2]+[C,H]$ provided $\beta^\parallel=\bar\beta_\parallel$. 
The remaining equations for the terms linear in the scalars are
\begin{equation}
\label{detT}
\Lambda \bar{\beta}_\perp \partial_{\varphi}(e^{i\varphi} \xi)= -i e^{i\varphi}c \bar{\alpha}\,,
\qquad
\Lambda \beta^\perp \partial_{\varphi}(e^{i\varphi} \xi)=-i e^{i\varphi}\xi c\alpha\,,
\end{equation}
which are simple algebraic relations on $\beta^\perp$, $\bar\beta_\perp$ and $c$.

In the generic case, we can have
\begin{equation}
\label{Pi=0loops}
\mathcal{L}= \begin{pmatrix}
A_{\varphi,I} +M_a{}^b r^a \bar{r}_b -\frac{i}{k} (\tilde{\mu}_{I\; \dot{1}}^{\; \dot{1}} - \tilde{\mu}_{I\; \dot{2}}^{\; \dot{2}}) & -i(\bar{\alpha} +\xi \Lambda \bar{\beta}_{\perp}) \psi_{\dot{1}-} -i\Lambda \bar{\beta}_{\perp} \psi_{\dot{2}+}\\
i (\alpha +\Lambda \beta^{\perp})\bar{\psi}_+^{\dot{1}}-i \xi\Lambda \beta^{\perp} \bar{\psi}_-^{\dot{2}} & A_{\varphi,I+1} +M_a{}^b \bar{r}_b r^a -\frac{i}{k} (\tilde{\mu}_{I+1\dot{1}}^{\quad\;\; \;\dot{1}} -\tilde{\mu}_{I+1\dot{2}}^{\quad\;\; \;\dot{2}})+c- \frac{1}{2}
\end{pmatrix},
\end{equation}
where
\begin{equation}
\label{Pi0M}
M_a{}^b= \begin{pmatrix}
0 & \frac{i}{k\Lambda} +\xi\alpha \bar{\beta}_{\perp}\\
\frac{i}{k\Lambda}+\bar{\alpha} \beta^{\perp} & \beta^{\parallel}
\end{pmatrix},
\end{equation}
with $a,b=\perp,\parallel$. Plugging in the solutions of \eqref{detT}, the resulting loops generically preserve 
only one supercharge. However, at some special points we find supersymmetry enhancement. In 
fact, we find some very interesting subclasses of those loops, which are 
analyzed in detail in Section~\ref{sec:casesPi=0}.

%%%%%%%%%%

\subsubsection{The special cases: $\xi=0$ and $\xi=\infty$}
\label{sec:2-node-xi}

Two further degenerations of the $\Pi=0$ supercharges are when $\xi$ in \eqref{xipar} vanishes or 
is infinite. Both cases are equivalent under the replacement of $\eta$ with $\bar\eta$ 
(or $Q^{\dot 2a+}_{\imath}$ and $Q^{\dot 1a-}_{\bar{\imath}}$) and for simplicity 
we focus on $\xi=0$. 
This means that the supercharge $Q$ is comprised 
of only the four supercharges $Q^{\dot 1a -}_{\bar{\imath}}$ and is nilpotent 
$Q^2=0$.

In all cases when $\Pi=0$, we have two scalar fields $r^\parallel$ and $\bar r_\parallel$ in 
\eqref{rparallel} that are annihilated by $Q$. For $\xi=0$, as can be seen from \eqref{Pi0Qr}, 
there are also two fermionic field in the hypermultiplet annihilated by $Q$. Those are 
$\psi_{\dot 2+}$ and $\bar\psi^{\dot1}_+$ and we can therefore insert any distribution 
of these fields in the hyperloop while still preserving supersymmetry.

As the bottom left entry in $\cL_{1/2}$ is comprised of $\bar\psi^{\dot1}_+$, see \eqref{L1/2}, 
the matrix $H$ appearing in the variation of $\cL_{1/2}$ is upper-triangular, as can indeed 
be read off from \eqref{Pi0H}. To construct the deformed loops we take $G$ as in \eqref{Pi=0G} 
and add the extra fermionic fields to $C$. Alternatively, they can also be added as extra terms 
into $F$ beyond $QG$
\begin{equation}
G=\begin{pmatrix}
0 & \bar{\beta}_\perp r^{\perp}\\
\beta^\perp \bar{r}_{\perp} & 0
\end{pmatrix},
\qquad
C=\begin{pmatrix}
\bar{\beta}_\parallel r^{\parallel}\bar r_{\parallel}&\bar\delta\psi_{\dot 2+}\\
\delta\bar\psi^{\dot1}_+&\beta^\parallel \bar{r}_{\parallel}r^\parallel+c
\end{pmatrix}.
\end{equation}

Plugging $G$ and $C$ into $Q \mathcal{L}=\mathcal{D}_{\varphi}^{\mathcal{L}}H$, one gets again the same condition that appeared in \eqref{pi=0susycondition}, which can be solved by
\begin{equation}
\delta=c=0\,,\qquad \bar{\beta}_{\parallel}= \beta^{\parallel}\,.
\end{equation}
This gives the superconnection
\begin{equation}
\label{xi=0}
\mathcal{L}= \begin{pmatrix}
A_{\varphi,I} +M_a{}^b r^a \bar{r}_b -\frac{i}{k} (\tilde{\mu}_{I\; \dot{1}}^{\; \dot{1}} - \tilde{\mu}_{I\; \dot{2}}^{\; \dot{2}}) & -i\bar{\alpha} \psi_{\dot{1}-} +(\bar{\delta}-i\Lambda \bar{\beta}_{\perp}) \psi_{\dot{2}+}\\
i (\alpha +\Lambda \beta^{\perp})\bar{\psi}_+^{\dot{1}} & A_{\varphi,I+1} +M_a{}^b \bar{r}_b r^a -\frac{i}{k} (\tilde{\mu}_{I+1\dot{1}}^{\quad\;\; \dot{1}} -\tilde{\mu}_{I+1\dot{2}}^{\quad\;\; \dot{2}})- \frac{1}{2}
\end{pmatrix},
\end{equation}
where $M_a{}^b$ is the same as \eqref{Pi0M} with $\xi=0$. Note that $\bar{\delta}$ and $\bar{\beta}_{\perp}$ appear 
only as the combination $\bar{\delta}-i\Lambda \bar{\beta}_{\perp}$, so we can eliminate any one of them.

The same answer is found from a different approach in Appendix~\ref{app:modification}, where extra fermionic fields are added in $F$.

%%%%%%%%%%%%%%%%%

\subsubsection{The condition $\Pi = 0$ from an algebraic point of view}
\label{sec:alg2}

As done for $\Pi\neq 0$ in Section~\ref{sec:alg1}, one can consider the condition $\Pi = 0$ from an algebraic point of view. Here we give a complete classification of all possible subcases. From the discussion around \eqref{lastcase}, with $Q^2 \in \mathfrak{u}(1)_{J_0} \oplus \su(2)_L \oplus \mathfrak{u}(1)_{L_\perp}$, the conditions on $\bar\eta^\imath_a, \eta^\imath_a$ for $\Pi$ to vanish are easily derived, since one only needs to enforce $\Pi_0=0$, so that $Q^2 \in \su(2)_L \oplus \mathfrak{u}(1)_{L_\perp}$. By~\eqref{PiLambdaNonconformal}, there are two possibilities: either $\epsilon^{ab} z_a w_b=0$ which implies $\lambda_{ab} = \lambda_{ba}$ and $ Q^2 \in \su(2)_L$, or $\epsilon_{\imath\jmath} \bar t^\imath t^\jmath =0$, which implies $\lambda_{ab} = -\lambda_{ba}$ and $Q^2 \in \mathfrak{u}(1)_{L_\perp}.$

In the former case, $Q^2 \in \su(2)_L$, one can let $z_a = w_a$ without loss of generality, leading to
\begin{align}
Q^2 \propto w_a w_b R^{ab}\,.
\end{align}
The functions $\xi$ and $\Lambda$ are given by 
\bal
\xi = \frac{t^l + e^{-i\varphi}t^r}{e^{+i\varphi} \bar t^l + \bar t^r}\,, 
\qquad 
\Lambda = (e^{+i\varphi} \bar t^l + \bar t^r)^2\,.
\eal
In the case $Q^2 \in \mathfrak{u}(1)_{L_\perp}$, one may write instead $t^\imath = t \, s^\imath, \bar t^\imath = \bar t \, s^\imath$, leading to
\begin{align}
Q^2 \propto \epsilon^{ab} z_a w_b \, L_\perp\,,
\end{align}
as well as to
\bal
\label{eqn:Q2U(1)xi}
\xi = \frac{t e^{-i\varphi}}{\bar t}\,, 
\qquad 
\Lambda = {\bar t} ^2 e^{i\varphi} \Big( e^{i\varphi} (s^l)^2 + e^{-i\varphi}(s^r)^2 + 2 s^l s^r (\epsilon^{ab} z_a w_b)\Big)\,.
\eal

Finally, when both of the conditions above are met the supercharge becomes nilpotent, $Q^2=0$. The parameters are of the form
\begin{align}
\label{nilpotent}
\eta^\imath_a = a \rho^\imath w_a\,,
\qquad 
\bar\eta^\imath_a = \bar a \rho^\imath w_a\,.
\end{align}
This factorisation is expected since each term in~\eqref{QQ'} antisymmetrises over either $\imath, \jmath$ or $a, b$ (or both).
The functions $\xi$ and $\Lambda$ take the simple form 
\bal
\xi = \frac{a e^{-i\varphi}}{\bar a}\,, 
\qquad 
\Lambda = \bar a^2 (e^{i\varphi} \rho^l + \rho^r)^2\,.
\eal

Note that the function $\xi$ provides a handy way of distinguishing these cases. Concretely, $\partial_\varphi \left( e^{i\varphi}\xi\right) = 0$ 
if and only if $Q^2 \in \mathfrak{u}(1)_{L_\perp}$. $\xi$ vanishes identically if and only if $Q$ is composed entirely of barred supercharges.

%%%%%%%%%%%%%%%%%%%%%%%

\section{Longer quivers and twisted hypers}
\label{sec:longer}

All the constructions in Section~\ref{sec:2-node} involve only two nodes of the quiver. Here 
we turn to hyperloops coupling to more nodes. As a guiding example and starting point of the deformation, we consider the 1/2 BPS loop on two pairs of nodes, with undeformed superconnection given by 
\beq
\label{4nodeL1/2}
\cL_{1/2} = \begin{pmatrix}
\cA_{I} %+ \frac{1}{2}
& -i \bar\alpha_I \psi_{I, \dot{1} -} &0&0\\
i \alpha_I \bar \psi^{\dot{1}}_{I,+} 
&\cA_{I+1}-\frac12&0&0
\\
0&0&\cA_{I+2}-c
& -i \bar\alpha_{I+2} \psi_{I+2, \dot{1} -} 
\\
0&0&i \alpha_{I+2} \bar \psi^{\dot{1}}_{I+2,+} 
&
\cA_{I+3}-c -\frac{1}{2}
\end{pmatrix}.
\eeq
We introduce a constant shift $c$ between the two pairs of nodes representing the effect of a $U(N_{I+1})$ gauge freedom.
In this block-diagonal form, there is no restriction on $c$. The resulting Wilson loop is well defined 
with constant $\alpha_{I+2}$ and $\bar\alpha_{I+2}$ satisfying $\alpha_{I+2}\bar\alpha_{I+2}=2i/k$. We find 
(the supertrace sums lines with signs $+,-,+,-$)
\beq
W=\sTr\cP\exp i\oint\cL\,d\varphi=W_{(I,I+1)}+\exp\left(-i\oint c\,d\varphi\right)W_{(I+2,I+3)}\,.
\eeq
Clearly with this block-diagonal structure, we can take any linear combination of the two Wilson 
loops. 
Adding deformations by the hypermultiplets keeps the block-diagonal structure, so again it works with any $c$. 
As already noted in \cite{Drukker:2020dvr}, deformations by twisted hypermultiplets with $\tilde q^{\dot a}_{I+1}$ 
are more subtle and fix $c$.

The Wilson loop based on \eqref{4nodeL1/2} still satisfies 
$Q \cL_{1/2} =\mathcal{D}_{\varphi}^{\cL_{1/2}}H$, 
this time with
\begin{equation}
\label{4nodeH}
H= \begin{pmatrix}
0 & \bar\alpha_{I}r_I^2 & 0 & 0 \\
\alpha_I\bar{r}_{I\,2} & 0 & 0&0 \\
0 & 0 & 0 & \bar\alpha_{I+2} r_{I+2}^2 \\
0 & 0 & \alpha_{I+2}\bar{r}_{I+2\,2} & 0
\end{pmatrix}.
\end{equation}
It is now natural to rotate the fermions from the twisted hypermultiplets
\beq
\label{rho}
\tilde\rho^1_{-}=-( \eta\bar v)_a \tilde\psi^a_-\,,
\quad
\tilde\rho^2_{+}=( \bar\eta v)_a \tilde\psi^a_+
\,,\quad
\bar{\tilde\rho}_{1+}=\epsilon^{ab}( \bar\eta v)_a \bar{\tilde\psi}_{b+}\,,
\quad
\bar{\tilde\rho}_{2-}=\epsilon^{ab}( \eta\bar v)_a \bar{\tilde\psi}_{b-}\,,
\eeq
such that the supersymmetry transformations are
\beq
\label{Qtildeq}
Q \tilde q_{\dot 1}=\tilde\rho^2_{+}\,,
\quad
Q \tilde q_{\dot2}=\tilde\rho^1_{-}\,,
\quad
Q \bar{\tilde q}^{\dot1}=\bar{\tilde\rho}_{2-}\,,
\quad
Q \bar{\tilde q}^{\dot2}=\bar{\tilde\rho}_{1+}\,.
\end{equation}
The double variations are then
\bal
\label{Q2tildeq}
Q^2 \tilde q_{\dot 1}&= \Pi\big(i\partial_\varphi \tilde q_{\dot 1} + \cA_{I+1} \tilde q_{\dot 1} - \tilde q_{\dot 1} \cA_{I+2}\big) 
- \frac{2i}{k} (\bar{r}_2 r^2 \tilde q_{\dot 1}-\tilde q_{\dot 1} r^2\bar{r}_2) 
-\frac{1}{2}\epsilon^{ab}(\bar\eta v)_a(\eta\sigma^3\bar{v})_b \tilde q_{\dot 1}\,,
\\
Q^2 \tilde q_{\dot2}&= \Pi\big(i\partial_\varphi \tilde q_{\dot 2} + \cA_{I+1} \tilde q_{\dot 2} - \tilde q_{\dot 2} \cA_{I+2}\big) 
- \frac{2i}{k} (\bar{r}_2 r^2 \tilde q_{\dot 2}-\tilde q_{\dot 2} r^2\bar{r}_2) 
-\frac{1}{2}\epsilon^{ab} (\bar\eta\sigma^3 v)_a(\eta \bar{v})_b \tilde q_{\dot 2}\,,
\\
Q^2 \bar{\tilde q}^{\dot1}
&= \Pi\big(i\partial_\varphi \bar{\tilde q}^{\dot1} + \cA_{I+2}\bar{\tilde q}^{\dot1} - \bar{\tilde q}^{\dot1} \cA_{I+1}\big) 
- \frac{2i}{k} ( r^2 \bar{r}_2\bar{\tilde q}^{\dot1}-\bar{\tilde q}^{\dot1} \bar{r}_2 r^2) 
-\frac{1}{2}\epsilon^{ab}(\bar\eta\sigma^3 v)_a(\eta\bar{v})_b \bar{\tilde q}^{\dot1}\,,
\\
Q^2 \bar{\tilde q}^{\dot2}
&=\Pi\big(i\partial_\varphi \bar{\tilde q}^{\dot2} + \cA_{I+2} \bar{\tilde q}^{\dot2} - \bar{\tilde q}^{\dot2} \cA_{I+1}\big) 
- \frac{2i}{k} ( r^2 \bar{r}_2\bar{\tilde q}^{\dot2}-\bar{\tilde q}^{\dot2} \bar{r}_2 r^2) 
-\frac{1}{2}\epsilon^{ab}(\bar\eta v)_a (\eta\sigma^3 v)_b \bar{\tilde q}^{\dot2}\,.
\eal
Using \eqref{sigma3}, the linear terms above can rewritten as
\bal
\epsilon^{ab}(\bar\eta v)_a(\eta\sigma^3\bar v)_b =-i\partial_\varphi \Pi- \lambda
\,,\qquad
\epsilon^{ab}(\bar\eta\sigma^3 v)_a(\eta\bar v)_b = 
-i\partial_\varphi \Pi +\lambda
\,,
\eal
such that the double variations become
\bal
\label{Q2tildeq2}
Q^2 \tilde q_{\dot 1}&= \Pi\left(i\partial_\varphi \tilde q_{\dot 1} -\frac12\tilde q_{\dot 1}+\Gamma\tilde q_{\dot 1}
+ \cA_{I+1} \tilde q_{\dot 1} - \tilde q_{\dot 1} \cA_{I+2}\right) 
- \frac{2i}{k} (\bar{r}_2 r^2 \tilde q_{\dot 1}-\tilde q_{\dot 1} r^2\bar{r}_2)\,,
%-\frac{1}{2}\Pi \tilde q_{\dot 1}
%+i\epsilon^{ab}(\bar\eta v)_a\partial_\varphi(\eta\bar{v})_b \tilde q_{\dot 1}\,,
\\
Q^2 \tilde q_{\dot2}
&= \Pi\left(i\partial_\varphi \tilde q_{\dot 2} +\frac{1}{2} \tilde q_{\dot 2}+\bar\Gamma\tilde q_{\dot 2}
+ \cA_{I+1} \tilde q_{\dot 2} - \tilde q_{\dot 2} \cA_{I+2}\right) 
- \frac{2i}{k} (\bar{r}_2 r^2 \tilde q_{\dot 2}-\tilde q_{\dot 2} r^2\bar{r}_2)\,,
%+\frac{1}{2}\Pi \tilde q_{\dot 2}
%+i\epsilon^{ab} \partial_\varphi(\bar\eta v)_a(\eta \bar{v})_b \tilde q_{\dot 2}\,,
\\
Q^2 \bar{\tilde q}^{\dot1}
&= \Pi\left(i\partial_\varphi \bar{\tilde q}^{\dot1} +\frac12\bar{\tilde q}^{\dot1}+\bar\Gamma\bar{\tilde q}^{\dot1}
+ \cA_{I+2}\bar{\tilde q}^{\dot1} - \bar{\tilde q}^{\dot1} \cA_{I+1}\right) 
- \frac{2i}{k} ( r^2 \bar{r}_2\bar{\tilde q}^{\dot1}-\bar{\tilde q}^{\dot1} \bar{r}_2 r^2) \,,
%+i\epsilon^{ab}\partial_\varphi(\bar\eta v)_a(\eta\bar{v})_b \bar{\tilde q}^{\dot1}\,,
\\
Q^2 \bar{\tilde q}^{\dot2}
&=\Pi\left(i\partial_\varphi \bar{\tilde q}^{\dot2} -\frac12\bar{\tilde q}^{\dot2}+\Gamma\bar{\tilde q}^{\dot2}
+ \cA_{I+2} \bar{\tilde q}^{\dot2} - \bar{\tilde q}^{\dot2} \cA_{I+1}\right) 
- \frac{2i}{k} ( r^2 \bar{r}_2\bar{\tilde q}^{\dot2}-\bar{\tilde q}^{\dot2} \bar{r}_2 r^2) \,,
%+i\epsilon^{ab}(\bar\eta v)_a\partial_\varphi (\eta v)_b \bar{\tilde q}^{\dot2}\,.
\eal
where for latter convenience we introduce
\beq
\Gamma = \frac{1}{2}\left(i\partial_\varphi\ln\Pi+\frac{\lambda}{\Pi}+1\right),
\qquad
\bar\Gamma = \frac{1}{2}\left(i\partial_\varphi\ln\Pi-\frac{\lambda}{\Pi}-1\right).
\eeq

%%%%%%%%%%%%%%%%%

\subsection{Deformations with $\Pi\neq0$}

We now proceed to deform the loop \eqref{4nodeL1/2} as in \eqref{L}. We take $G$ to be of the form
\begin{equation}
\label{G4node}
G = \begin{pmatrix}
0 & \bar{b}_{Ia} r_I^a & 0 & 0 \\
b_I^a \bar{r}_{Ia} & 0 & \bar{d}^{\dot 1}_{I+1}\tilde{q}_{I+1\,\dot 1} & 0 \\
0 & {d}_{I+1\,\dot 1}\bar{\tilde q}^{\dot 1}_{I+1} & 0 & \bar{b}_{I+2\,a} r_{I+2}^a \\
0 & 0 & b_{I+2}^a \bar{r}_{I+2\,a} & 0
\end{pmatrix}.
\end{equation}
We allow a coupling to all the scalars in the hypermultiplets, but in the twisted hypers we restrict
to $\tilde{q}_{I+1\,\dot 1}$ and $\bar{\tilde q}^{\dot 1}_{I+1}$. The second pair of scalar fields is examined below.

Using \eqref{Q2tildeq2}, the analogue of \eqref{Q2G} adapted for a longer quiver is
\begin{equation}
\label{Q2G-4nodes}
-iQ^2 G= \partial_\varphi(\Pi G)- i [\cL_{1/2}^B,\Pi G] + i[H^2,G] - \Pi\hat{\cG}\,,
\end{equation}
with
\begin{equation}
\begin{aligned}
\Pi\hat{\cG}&= \begin{pmatrix}
0 & \partial_\varphi(\Pi\bar b_{I\,a}) r_I^a & 0 & 0 \\
\partial_\varphi(\Pi b_I^a) \bar r_{I\,a} & 0 & \partial_\varphi(\Pi\bar{d}^{\dot 1})\tilde{q}_{I+1\,\dot 1} & 0 \\
0 & \partial_\varphi(\Pi{d}_{\dot 1})\bar{\tilde q}_{I+1}^{\dot 1} & 0 & \partial_\varphi(\Pi\bar b_{I+2\,a}) r_{I+2}^a \\
0 & 0 & \partial_\varphi(\Pi b_{I+2}^a) \bar r_{I+2\,a} & 0
\end{pmatrix}\\
&\quad+\begin{pmatrix}
0 & -i\Pi \bar{b}_{I\,1} r^1_I & 0 & 0 \\
i\Pi b_I^1 \bar r_{I\,1} & 0 & -i\Pi(c-\Gamma)\bar d^{\dot 1}\tilde q_{I+1\,\dot 1} & 0 \\
0 & i\Pi(c+\bar\Gamma)d_{\dot 1}\bar{\tilde q}^{\dot1}_{I+1} & 0 & -i\Pi \bar{b}_{I+2\,1} r_{I+2}^1 \\
0 & 0 & i\Pi b^1_{I+2} \bar r_{I+2\,1} & 0
\end{pmatrix}.
\end{aligned}
\end{equation}

Proceeding as before, the analogue of \eqref{QcL} sets $B=\{G,H\}+\Pi G^2$ and 
supersymmetry invariance of $\cL$ now requires solving
\beq
\label{constraintPiG}
i[C,H+\Pi G]-\Pi \hat{\cG} =0\,,
\qquad
C=\diag(c_I,c_{I+1},c_{I+2},c_{I+3})\,.
\eeq
We recover two copies of the equations \eqref{4diffeq}, now for $b_I$, $\bar b_I$, $b_{I+2}$, and $\bar b_{I+2}$. 
In addition, using $\Gamma+\bar\Gamma=i\partial_\varphi\ln\Pi$, 
we find the two following equations for $d_{I+1\,\dot 1}$ and $\bar d^{\dot1}_{I+1}$
\bal
\label{4diffeq-twisted}
\partial_{\varphi} (\bar{d}^{\dot 1}_{I+1}) -i(c_{I+1}-c_{I+2}+c+\bar\Gamma) \bar{d}_{I+1}^{\dot 1}&=0\,, \\
\partial_{\varphi} (\Pi d_{I+1\,\dot{1}}) +i (c_{I+1}-c_{I+2} +c+\bar\Gamma) \Pi {d}_{I+1\,\dot{1}} &=0\,.
\eal
Note that these involve not only the numerical factors arising from $C$ but also the relative shift $c$ that 
was left arbitrary in \eqref{4nodeL1/2}. In particular, we can make use of this gauge freedom to make the 
convenient choice $c_{I+1}=c_{I+2}=0$ and then with $c=-\bar\Gamma$, the 
equations above are solved by
\beq
\label{sol-d1}
\bar d^{\dot1}_{I+1}=\bar\delta^{\dot1}_{I+1}\,, 
\qquad 
d_{I+1\,\dot{1}}=\frac{\delta_{I+1\,\dot1}}{\Pi}\,,
\eeq
with constant $\delta$'s. Other gauges are possible, but they are completely equivalent to this one. 

One can write the explicit expression for $\cL$ using \eqref{deformed-L}. Two points to note are 
that in addition to the diagonal bosonic terms and first off-diagonal fermionic terms, there are also 
off-off-diagonal bosonic terms that contain the bilinears $\tilde q_{I+1\,\dot 1}r^a_{I+2}$ and 
$\bar{\tilde q}_{I+1}^{\dot 1}\bar r_{Ia}$. Also, the diagonal terms in the central nodes now include the modification 
of the bilinears of the scalars in the twisted hypermultiplets via
\beq
%\tilde\mu_{I+1}^{\dot 1}{}_{\dot1}-\tilde\mu_{I+1}^{\dot 2}{}_{\dot2}\to
\widetilde{M}^{\dot a}{}_{\dot b}\tilde q_{I+1\,\dot a}\bar{\tilde q}_{I+1}^{\dot b}\,,
\qquad
\widetilde{M}= \begin{pmatrix}
-i/k + {\bar\delta}^{\dot 1}_{I+1}\delta_{I+1\,\dot 1} & 0 \\
0 & i/k
\end{pmatrix}.
\eeq
Instead of writing the full complicated $4\times4$ form of the general $\cL$, we look at some special 
cases in Section~\ref{sec:cases-twisted}.

To couple $\cL$ to $\tilde{q}_{I+1\,\dot 2}$ and $\bar{\tilde{q}}^{\dot 2}_{I+1}$, we take instead
\begin{equation}
\label{G4node-2}
G = \begin{pmatrix}
0 & \bar{b}_{Ia} r_I^a & 0 & 0 \\
b_I^a \bar{r}_{Ia} & 0 & \bar{d}^{\dot 2}_{I+1}\tilde{q}_{I+1\,\dot 2} & 0 \\
0 & {d}_{I+1\,\dot 2}\bar{\tilde q}^{\dot 2}_{I+1} & 0 & \bar{b}_{I+2\,a} r_{I+2}^a \\
0 & 0 & b_{I+2}^a \bar{r}_{I+2\,a} & 0
\end{pmatrix},
\end{equation}
then \eqref{Q2G-4nodes} holds with
\begin{equation}
\begin{aligned}
\Pi\hat{\cG}&= \begin{pmatrix}
0 & \partial_\varphi(\Pi\bar b_{I\,a}) r_I^a & 0 & 0 \\
\partial_\varphi(\Pi b_I^a) \bar r_{I\,a} & 0 & \partial_\varphi(\Pi\bar{d}^{\dot 2}_{I+1})\tilde{q}_{I+1\,\dot{2}} & 0 \\
0 & \partial_\varphi(\Pi d_{I+1\,\dot 2})\bar{\tilde{q}}^{\dot 2}_{I+1} & 0 & \partial_\varphi(\Pi\bar b_{I+2\,a}) r_{I+2}^a \\
0 & 0 & \partial_\varphi(\Pi b_{I+2}^a) \bar r_{I+2\,a} & 0
\end{pmatrix}\\
&\quad+\begin{pmatrix}
0 & -i\Pi \bar{b}_{I\,1} r^1_I & 0 & 0 \\
i\Pi b_I^1 \bar r_{I\,1} & 0 & -i\Pi(c-\bar\Gamma-1)\bar{d}^{\dot 2}_{I+1}\tilde{q}_{I+1\,\dot{2}} & 0 \\
0 & i\Pi(c+\Gamma-1)d_{I+1\,\dot 2}\bar{\tilde{q}}^{\dot 2}_{I+1} & 0 & -i\Pi \bar{b}_{I+2\,1} r_{I+2}^1 \\
0 & 0 & i\Pi b^1_{I+2} \bar r_{I+2\,1} & 0
\end{pmatrix}\,.
\end{aligned}
\end{equation}
This time, \eqref{constraintPiG} gives two equations for $d_{I+1\,\dot 2}$ and $\bar{d}^{\dot 2}_{I+1}$
\bal
\label{4diffeq-twisted2}
\partial_{\varphi} (\bar{d}^{\dot 2}_{I+1}) -i(c_{I+1}-c_{I+2}+c+\Gamma-1) \Pi\bar{d}^{\dot 2}_{I+1}&=0\,, \\
\partial_{\varphi} (\Pi d_{I+1\,\dot{2}}) +i (c_{I+1}-c_{I+2} +c+\Gamma -1) {d}_{I+1\,\dot{2}} &=0\,.
\eal
In this case the convenient gauge is $c_{I+1}=c_{I+2}=0$ where these equations are solved with $c=-\Gamma+1$ and
\beq
\label{sol-d2}
\bar{d}^{\dot 2}_{I+1}= \bar\delta^{\dot 2}_{I+1}\,,\qquad d_{I+1\,\dot 2}= \frac{\delta_{I+1\,\dot 2}}{\Pi},
\eeq
with constant $\delta$'s.
Now $\widetilde{M}$ is given by
\beq
\widetilde{M}= \begin{pmatrix}
-i/k & 0 \\
0 & i/k + {\bar\delta}^{\dot 2}_{I+1}\delta_{I+1\,\dot 2}
\end{pmatrix}.
\eeq

Notice that we performed the analysis separately for the two pairs of scalars in the twisted hypermultiplets and 
the resulting expressions required different conditions on $c$, namely $c=-\bar\Gamma$ and $c=-\Gamma+1$. 
To allow $\cL$ to couple to all scalars of the twisted hypermultiplet at the same time, these need to be related 
by a gauge transformation, requiring
\beq
\hat c(\varphi)=- \int_0^\varphi \left( \bar\Gamma - \Gamma + 1 \right) d\varphi' = \int_0^\varphi \frac{\lambda}{\Pi}\, d\varphi'\,,
\eeq
to be single valued. Thus, if
\beq
\label{obstruction}
e^{i\hat c(2\pi)}=\exp i\oint \frac{\lambda}{\Pi}\, d\varphi=1\,,
\eeq
is satisfied, $\cL$ may couple to all twisted scalars, otherwise it may couple either to the pair 
$\tilde{q}_{\dot 1},\bar{\tilde{q}}^{\dot 1}$ or to $\tilde{q}_{\dot 2},\bar{\tilde{q}}^{\dot 2}$.
 
To be concrete, if we choose the gauge $c_I=c_{I+1}=c_{I+2}=c_{I+3}=0$ and $c=-\bar\Gamma$, 
a $G$ including all twisted scalars is then composed from \eqref{G4node} and the gauge transformed 
version of \eqref{G4node-2}, giving
\begin{equation}
G = \begin{pmatrix}
0 & \bar{b}_{Ia} r_I^a & 0 & 0 \\
b_I^a \bar{r}_{Ia} & 0 & 
\bar{d}^{\dot 1}_{I+1}\tilde{q}_{I+1\,\dot 1} + e^{i\hat c(\varphi)}\bar{d}^{\dot 2}_{I+1}\tilde{q}_{I+1\,\dot 2} & 0 \\
0 & {d}_{I+1\,\dot 1}\bar{\tilde q}^{\dot 1}_{I+1} +e^{-i\hat c(\varphi)}{d}_{I+1\,\dot 2}\bar{\tilde q}^{\dot 2}_{I+1} 
& 0 & \bar{b}_{I+2\,a} r_{I+2}^a \\
0 & 0 & b_{I+2}^a \bar{r}_{I+2\,a} & 0
\end{pmatrix}.
\end{equation}
The construction then follows as before. Differential equations for 
$\bar{b}_{Ia}$, $b_I^a$, $\bar{b}_{I+2\,a}$, $b_{I+2}^a$ and for $d_{I+1\,\dot 1}$, $\bar{d}_{I+1}^{\dot1}$ 
are as in \eqref{4diffeq} and \eqref{4diffeq-twisted} and are solved by \eqref{beta} and \eqref{sol-d1}. 
As for $d_{I+1\,\dot 2},\bar{d}_{I+1}^{\dot 2}$, we find the equivalent to \eqref{4diffeq-twisted2} in the $c=-\bar\Gamma$ gauge
\bal
\partial_{\varphi} (\Pi e^{i\hat c(\varphi)} \bar{d}^{\dot 2}_{I+1}) -i\lambda e^{i\hat c(\varphi)} \bar{d}^{\dot 2}_{I+1}&=0\,, \\
\partial_{\varphi} (\Pi e^{-i\hat c(\varphi)} d_{I+1\,\dot{2}}) +i \lambda e^{-i\hat c(\varphi)}{d}_{I+1\,\dot{2}} &=0\,,
\eal
which is still solved by \eqref{sol-d2}.

We found therefore the form of $\cL$ coupling to both twisted scalars, under the condition \eqref{obstruction}. Now $\widetilde{M}$ is given by
\beq
\widetilde{M}= \begin{pmatrix}
-i/k + {\bar\delta}^{\dot 1}_{I+1}\delta_{I+1\,\dot 1} & e^{-i\hat c(\varphi)} {\bar\delta}^{\dot 1}_{I+1}\delta_{I+1\,\dot 2} \\
e^{i\hat c(\varphi)} {\bar\delta}^{\dot 2}_{I+1}\delta_{I+1\,\dot 1} & i/k + {\bar\delta}^{\dot 2}_{I+1}\delta_{I+1\,\dot 2}
\end{pmatrix}.
\eeq
A special case of this construction was already carried out in \cite{Drukker:2020dvr}. In the parameterization of that paper, $\Pi=1$ and $\lambda=\cos\theta$, with $\theta$ the so-called ``latitude'' angle. It was then possible to include all scalar fields in $G$ for $\theta=0$ (see equation (4.9) of \cite{Drukker:2020dvr}). 
The analog of the obstruction \eqref{obstruction} arose there for $\theta\neq0$ (see the comment below (5.15) 
of \cite{Drukker:2020dvr}). The reasoning for that is precisely the fact that 
$e^{i\hat c(\varphi)}=e^{i\varphi\cos\theta}$ considered there is not single valued.

%%%%%%%%%%%%%%%%%%%%%%%%%%%%%

\subsection{Deformations with $\Pi=0$}
\label{sec:4node-0}

Generalizing Section~\ref{sec:2-node-0} to allow for twisted hypers, we start again with the $1/2$ BPS loop 
with four nodes in \eqref{4nodeL1/2}. $H$ is the same as in \eqref{4nodeH}, now written generalizing \eqref{Pi0H} to
\beq
H= \begin{pmatrix}
0 & \bar\alpha_I r_I^\parallel & 0 & 0 \\
-\alpha_I \xi\bar{r}_{I\,\parallel} & 0 & 0 & 0\\
0 & 0 & 0 & \bar\alpha_{I+2}r^\parallel_{I+2} \\
0 & 0 & -\alpha_{I+2}\xi\bar{r}_{I+2\,\parallel} & 0
\end{pmatrix}.
\eeq
As before, the fact that $\Pi=0$ implies that the variation of the deformed loop is still a covariant derivative of $H$ 
regardless of the deformation. Since $H$ does not include twisted scalars, we do not expect the relative shift between 
the two pairs of nodes ($c$ in \eqref{4nodeL1/2}) to be fixed by the requirement that the deformed loop is 
supersymmetric. Below we see that this is indeed the case.

The $\Pi=0$ version of the double transformations \eqref{Q2tildeq} is 
\bal
\label{Q2tildeq2-Pi0}
Q^2 \tilde q_{\dot 1}&=
 \frac{2i}{k} \xi (\bar{r}_\parallel r^\parallel \tilde q_{\dot 1}-\tilde q_{\dot 1} r^\parallel\bar{r}_\parallel) + \frac{\lambda}{2} \tilde{q}_{\dot 1}\,,
\\
Q^2 \tilde q_{\dot2}
&= \frac{2i}{k} \xi (\bar{r}_\parallel r^\parallel \tilde q_{\dot 2}-\tilde q_{\dot 2} r^\parallel\bar{r}_\parallel) -\frac{\lambda}{2} \tilde{q}_{\dot 2}\,,
\\
Q^2 \bar{\tilde q}^{\dot1}
&= \frac{2i}{k} \xi( r^\parallel \bar{r}_\parallel\bar{\tilde q}^{\dot1}-\bar{\tilde q}^{\dot1} \bar{r}_\parallel r^\parallel) -\frac{\lambda}{2} \bar{\tilde q}^{\dot1}\,,
\\
Q^2 \bar{\tilde q}^{\dot2}
&= \frac{2i}{k} \xi( r^\parallel \bar{r}_\parallel\bar{\tilde q}^{\dot2}-\bar{\tilde q}^{\dot2} \bar{r}_\parallel r^\parallel) + \frac{\lambda}{2} \bar{\tilde q}^{\dot2}\,.
\eal
The building blocks are then the $4\times 4$ versions of $G$ and $C$ 
(we set $c_I=c_{I+2}=0$ for convenience)
\begin{equation}
\begin{aligned}
\label{4nodeGC-0}
G&=\begin{pmatrix}
0 & \bar\beta_{I\,\perp} r^\perp_I & 0 & 0 \\
\beta_I^\perp \bar{r}_{I\,\perp} & 0 & \bar{d}^{\dot a} \tilde{q}_{I+1\,\dot{a}} & 0 \\
0 & d_{\dot a}\bar{\tilde{q}}^{\dot a}_{I+1} & 0 & \bar\beta_{I+2\,\perp} r^\perp_{I+2} \\
0 & 0 & \beta_{I+2}^\perp \bar{r}_{I+2\perp} & 0
\end{pmatrix},
\\
C &= \begin{pmatrix}
\bar\beta_{I\,\parallel} r_I^\parallel \bar{r}_{I\,\parallel} & 0 & 0 & 0 \\
0 & \beta_I^\parallel \bar{r}_{I\,\parallel} r^{\parallel}_{I} + c_{I+1} & 0 & 0 \\
0 & 0 & \bar\beta_{I+2\,\parallel} r_{I+2}^\parallel \bar{r}_{I+2\,\parallel} & 0 \\
0 & 0& 0 & \beta_{I+2}^\parallel \bar{r}_{I+2\,\parallel} r^{\parallel}_{I+2} + c_{I+3}
\end{pmatrix}.
\end{aligned}
\end{equation}
With these in hand, the superconnection $\cL = \cL_{1/2} -i QG + \{G,H\} + C$
is supersymmetric provided that the same condition as in \eqref{pi=0susycondition} is obeyed. 

The analysis for the $\beta$ parameters follows as in the 2-node case. Cubic terms on the fields cancel for 
$\bar\beta_{I\,\parallel}=\beta_{I}^\parallel$ and $\bar\beta_{I+2\,\parallel}=\beta_{I+2}^\parallel$. 
Linear terms are such that we find, in addition to \eqref{detT}, its $I+2$-node version
\begin{equation}
\label{detT'}
\Lambda \bar{\beta}_{I+2\,\perp} \partial_{\varphi}(e^{i\varphi} \xi)= -i e^{i\varphi}c_{I+3} \bar{\alpha}_{I+2}\,,
\qquad
\Lambda \beta_{I+2}^\perp \partial_{\varphi}(e^{i\varphi}\xi)=i e^{i\varphi}c_{I+3}\alpha_{I+2} \xi\,.
\end{equation}

For the central block containing the $d$ parameters, one realizes that the cubic term in the double variations 
\eqref{Q2tildeq2-Pi0} is exactly equal to $[G,H^2]$. There is no contribution related to $d$ from $[C,H]$, so 
one is left with the linear terms arising from $Q^2G$
\beq
\begin{pmatrix}
\ddots & ~ & ~ & ~ \\
~ & 0 & \dfrac{\lambda}{2}(\bar{d}^{\dot 1}\tilde{q}_{\dot 1} -\bar{d}^{\dot 2}\tilde{q}_2) & ~ \\
~ & -\dfrac{\lambda}{2} (d_{\dot 1}\bar{\tilde{q}}^{\dot 1} -d_{\dot 2}\bar{\tilde{q}}^{\dot 2}) & 0 & ~ \\
~ & ~ & ~ & \ddots
\end{pmatrix}
=0\,.
\eeq
Solutions with nonvanishing $d$ parameters and a non block-diagonal structure are only possible for supercharges with
\beq
\label{extraset}
\lambda =0\,.
\eeq
In this case there are no constraints on $\bar{d}^{\dot a}$ and $d_{\dot a}$, and they can be arbitrary functions.
At the level of the algebra, see Section~\ref{sec:alg2}, this means that loops in this section are constructed 
from $Q$'s that square only to $\su(2)_L$ generators $R^{ab}$.

Note that as anticipated the derivation above does not set restrictions on the relative shift $c$ appearing in 
\eqref{4nodeL1/2}, in contrast to the $\Pi\neq 0$ case. We examine some special cases of the resulting operators in 
Section~\ref{sec:cases-twistedPi0}.

%%%%%%%%%%%%%

\subsubsection{The special cases: $\xi=0$ and $\xi=\infty$}
\label{sec:4-node-xi0}

The analysis of $\xi=0$ and $\xi=\infty$ follows in analogy with Section~\ref{sec:2-node-xi}. As before, both cases are equivalent under the replacement of $\eta$ and $\bar\eta$, so we focus only on the $\xi=0$ case.

Here, since we are considering longer quivers coupling to twisted hypermultiplets, we need to include in $G$ not only $r^\perp$, $\bar{r}_\perp$ but also the twisted scalars that are not annihilated by $Q$. From \eqref{Qtildeq} we see that these are $\tilde{q}_{\dot 1}$ and $\bar{\tilde{q}}^{\dot 2}$, so we have
\beq
G = \begin{pmatrix}
0 & \bar\beta_{I\perp} r_I^\perp & 0 & 0 \\
\beta_I^\perp & 0 & \bar{d}^{\dot 1} \tilde{q}_{I+1\,\dot{1}} & 0 \\
0 & d_{\dot 2} \bar{\tilde{q}}_{I+1}^{\dot 2} & 0 & \bar\beta_{I+2} r_{I+2}^\perp \\
0 & 0 & \beta_{I+2}^\perp \bar{r}_{I+2\perp} & 0
\end{pmatrix}.
\eeq

Conversely, the fields $\tilde{q}_{\dot 2}$ and $\bar{\tilde{q}}^{\dot 1}$ are annihilated by $Q$ and are included in the matrix $C$. In addition to them, we should also include $\tilde\rho_+^2$ and $\bar{\tilde{\rho}}_{1+}$, which are the linear combination of fermionic fields from the twisted hypermultiplet that are annihilated by $Q$. Thus, we have (setting again $c_I$ and $c_{I+2}$ to zero for convenience)
\bal
C =&
\begin{pmatrix}
\bar K_I & \bar\delta_I \psi_{I\dot{2}+} & \gamma_1 r_I^\parallel \tilde{q}_{I+1\,\dot{2}} & 0 \\
\delta_I \bar\psi_{I+}^{\dot 1} & K_I + c_{I+1} & \bar\delta_{I+1} \tilde\rho_{I+1,+}^2 & \gamma_2 \tilde{q}_{I+1\,\dot{2}} r_{I+2}^{\parallel} \\
\gamma_3 \bar{\tilde{q}}_{I+1}^{\dot 1} \bar{r}_{I\parallel} & \delta_{I+1} \bar{\tilde{\rho}}_{I+1,1+} & \bar K_{I+2} & \bar\delta_{I+2} \psi_{I+2\,\dot{2}+} \\
0 & \gamma_4 \bar{r}_{I+2\parallel}\bar{\tilde{q}}_{I+1}^{\dot 1} & \delta_{I+2} \bar\psi_{I+2,+}^{\dot 1} & K_{I+2} + c_{I+3}
\end{pmatrix},
\eal
with $K_I\equiv \beta_{I}^\parallel \bar{r}_{I\parallel}r_I^\parallel + \tau_{I+1} \tilde{q}_{I+1\,\dot{2}} \bar{\tilde{q}}^{\dot 1}_{I+1}$ and $\bar K_I \equiv \bar\beta_{I\parallel} r_I^\parallel \bar{r}_{I\parallel} + \tau_I \bar{\tilde{q}}_{I-1}^{\dot 1} \tilde{q}_{I-1\,\dot{2}}$.
 
As before, the superconnection $\cL = \cL_{1/2} - iQG + \{G,H\} + C$ is supersymmetric provided that \eqref{pi=0susycondition} is obeyed. This is solved by
\beq
\bar\beta_{I\parallel} = \beta_I^\parallel \,, 
\qquad 
\bar\beta_{I+2\parallel} = \beta_{I+2}^\parallel\, , 
\qquad 
\gamma_1 \bar\alpha_{I+2}= \gamma_2 \bar\alpha_I \,,
\eeq
and by setting the remaining parameters in $C$ to zero, except for $\bar{\delta}_I$, 
$\bar\delta_{I+2}$ and $\delta_{I+1}$, which are left arbitrary. We write down the resulting operator at the end of 
Section~\ref{sec:cases-twistedPi0}.

%%%%%%%%%%%%%%%%%%%%

\section{Special cases}
\label{sec:cases}

Having carried out the systematic construction of BPS hyperloops described above, we turn now to some special examples of the constructions. This includes making 
contact with previously described operators and identifying new ones. Our emphasis is on operators preserving more than one supercharge.

\subsection{Single node bosonic loops}
\label{sec:bosonic}

We start with the simplest possible BPS Wilson loops in three-dimensional Chern-Simons-matter theories, those 
involving only a single node and $\cL$ is a $1\times1$ block with only the gauge field and bilinears of the scalars. 
The first such bosonic loops were constructed by Gaiotto and Yin 
in off-shell $\cN=2$ language in \cite{Gaiotto:2007qi}. Analogues of them in ABJ(M) theory were described in 
\cite{Drukker:2008zx, Chen:2008bp, Rey:2008bh} and that description carries over also 
to $\cN=4$ theories. Such loops preserve at most four supercharges. 
The other previously identified family of bosonic loops are the ``bosonic latitude'' loops of 
\cite{Cardinali:2012ru,Bianchi:2018bke,Drukker:2020dvr}, which preserve a pair of supercharges.

To get such loops in our setting we may decouple the nodes by simply setting 
$\beta^1=\beta^2=\bar\beta_1=\bar\beta_2=0$ in the analysis of Section~\ref{sec:2-node} for the case 
$\Pi\neq 0$ (we comment below on the case $\Pi=0$). This eliminates all the fermions in the 
superconnection $\cL$, which becomes block-diagonal with a connection in the $I$-th block taking the form
\bal
\label{Mrr}
\cA &= A_\varphi + \frac{i}{k}\Pi^{-1}(r^1\bar r_1-r^2\bar r_2) -\frac{i}{k}(\tilde \mu\indices{^{\dot{1}}_{\dot{1}}}
-\tilde \mu\indices{^{\dot{2}}_{\dot{2}}})
%=A_\varphi + \frac{i}{k}q^a M\indices{_a^b} \bar q_b 
%- \frac{i}{k} {\bar{\tilde q}}^{\dot{a}} \left(\sigma_3 \right)\indices{_{\dot{a}}^{\dot{b}}} {\tilde q}_{\dot{b}} 
\,,
\eal

It is easy to show that these loops preserve at least two supercharges. 
Consider in fact the supercharge $Q'$ gotten by the replacement $\bar\eta^\imath_a\to-\bar\eta^\imath_a$ in \eqref{Q}
\beq
Q^{\prime}
 = \eta^\imath_a Q^{\dot 2a+}_{\imath} 
- \bar\eta^\imath_a (\sigma^1)_\imath^{\ \bar\imath} Q^{\dot1a-}_{\bar\imath}\,.
\eeq
Under this change of sign, $\Pi\to-\Pi$, $r^2\to-r^2$ and $\bar r_1\to-\bar r_1$, such that \eqref{Mrr} 
is left invariant. Note that because $\Pi\neq0$, $Q$ is the sum of barred and unbarred supercharges and by the above argument these must be preserved separately.

Alternatively, this can be seen by investigating the bosonic symmetries. In particular, note that the transverse 
rotation $T_\perp$ keeps the loop fixed pointwise, and therefore acts trivially on the scalars as well as on the 
parallel component of the gauge field, the only fields in the bosonic loop. 
Closure of the symmetry algebra then implies that, in addition to $Q$, the supercharge 
$[T_\perp, Q]$ is preserved by the loop. From \eqref{Tperp} we see that this generates $Q^\prime$, 
so we come to the same conclusion as above 
(an analogous argument can be made using the generator $\bar R_3$).

A useful way to write the connection \eqref{Mrr} is in terms of the moment maps $\mu^a{}_b$ as
\beq
\label{randombosonic}
\cA = A_\varphi + \frac{i}{k}\frac{1}{(\chi-\bar\chi)}\left((\chi+\bar\chi)(\mu\indices{^{1}_{1}}
- \mu\indices{^{2}_{2}}) +2\mu\indices{^{2}_{1}}- 2\chi\bar\chi\, \mu\indices{^{1}_{2}}\right) 
-\frac{i}{k}(\tilde \mu\indices{^{\dot{1}}_{\dot{1}}}
-\tilde \mu\indices{^{\dot{2}}_{\dot{2}}})\,,
\eeq
with
\beq
\chi =\frac{(\eta \bar v)_1}{(\eta \bar v)_2}\,,
\qquad
\bar\chi = \frac{(\bar\eta v)_1}{(\bar\eta v)_2}\,,
\eeq
which are generally linear fractional transformations of $e^{i\varphi}$ \eqref{vs} (and as usual, they are not conjugates).

The most degenerate case is when both $\chi$ and $\bar\chi$ have no $\varphi$ dependence. This requires the 
numerators and denominators to be proportional to each-other, spanning a two dimensional space of $\eta$'s and 
likewise $\bar\eta$'s. This implies that the loop preserves 4 supercharges and having no $\varphi$ dependence, it 
also preserves the $SO(2,1)$ conformal group. To recover the Gaiotto-Yin Loop \cite{Gaiotto:2007qi} we take 
$\chi=1/\bar\chi\to\infty$. Other values of $\chi$, $\bar\chi$ are related by the action of the complexification of the 
broken $SU(2)_L$ symmetry.

When $\chi$ is a constant and $\bar\chi$ depends on $\varphi$ (or vice versa), there is only partial degeneracy, 
and the loops preserve three supercharges, or are 3/16 BPS. Such loops have not been previously discussed 
in the literature.

When both $\chi$ and $\bar\chi$ depend on $\varphi$, the loops preserve a pair of supercharges. A simple 
example is when they are just monomials, for example
$\chi=-\tan(\theta/2)e^{-i\varphi}$ and $\bar\chi=\cot(\theta/2) e^{-i\varphi}$. The connection takes the form 
\beq
\cA = A_\varphi -\frac{i}{k} \left( \cos\theta(\mu\indices{^{1}_{1}}
- \mu\indices{^{2}_{2}}) + \sin\theta\, e^{-i\varphi}\mu\indices{^{1}_{2}}
+ \sin\theta\, e^{i\varphi}\mu\indices{^{2}_{1}}\right) -\frac{i}{k}(\tilde \mu\indices{^{\dot{1}}_{\dot{1}}}
-\tilde \mu\indices{^{\dot{2}}_{\dot{2}}})\,.
\eeq
These are the latitude loops found in \cite{Cardinali:2012ru} and studied in \cite{Bianchi:2018bke,Drukker:2020dvr}. 
As the $\varphi$ dependence breaks conformal invariance, acting with the (complexified) conformal group 
$SL_2(\bC)$ on the loop above generates many other loops, including those where $\chi$ and $\bar\chi$ are 
proper rational functions and not mere monomials.

There are yet more peculiar bosonic loops that preserve two supercharges, but are not similar to the latitude loops. 
Representatives of those have
\beq
\chi=e^{-i\varphi}+\nu\,,
\qquad
\bar\chi=e^{-i\varphi}-\nu\,,
\eeq
with an arbitrary parameter $\nu$.

Despite all the machinery in the previous sections, the analysis of the most general BPS bosonic loop requires 
yet further techniques, so those will be explored in a future publication \cite{toappear}. That exploration will also 
relax the condition in this paper that the loops arise from continuous deformations of the 1/2 BPS loop, which 
could give rise to further BPS bosonic loops.

%%%%%%%%%%%%%%%%%%%%%%%

\subsection{Two-node hyperloops with $\Pi\neq0$}
\label{sec:cases2nodes}

Let us look now at some special examples of the hyperloops with two nodes constructed in 
Section~\ref{sec:2-node-not0}. Examining \eqref{beta-M}, the most symmetric possibility is 
that $M$ is proportional to the identity, restoring $SU(2)_L$ symmetry. 
There are two such solutions. The first with $\beta^1=\bar\beta_1=0$ and $\beta^2\bar\beta_2=2i/k$, 
which is just the original 1/2 BPS loop in \eqref{L1/2}. The second has 
$\beta^2=\bar\beta_2=0$ and $\beta^1\bar\beta_1=-2i/k$, which is the second 1/2 BPS loop with the 
same symmetries in \eqref{L1/2prime} (albeit written in a different gauge).

A less symmetric case is when $M$ is diagonal, but not necessarily proportional to the identity, so when 
$\bar{\beta}_1 \beta^{2} =\bar{\beta}_2 \beta^{1}=0$. If $\beta^1=\beta^2=0$ or 
$\bar{\beta}_1=\bar{\beta}_2=0$, the connection becomes upper or lower triangular, respectively. As discussed in \cite{Drukker:2019bev,drukker2020bps, Drukker:2020dvr}, 
the resulting loops are effectively the same as if all the $\beta^a=\bar\beta_a=0$, since they are all 
identical as quantum operators. The interesting case is 
then when $\bar{\beta}_1=\beta^{1}=0$ or $\bar{\beta}_2=\beta^{2}=0$. Taking the former as 
an example, we find
\begin{equation}
\cL=\begin{pmatrix}
A_{\varphi,I} + M_a^{\ b} r^a \bar{r}_b 
-\frac{i}{k} (\tilde{\mu}_{I \ \dot{1}}^{\ \dot{1}} - \tilde{\mu}_{I \ \dot{2}}^{\ \dot{2}}) 
& -i \bar{\beta}_2 \psi_{I \dot{1}-} \\
i\beta^{2} \bar{\psi}_{I+}^{\dot{1}} 
& A_{\varphi, I+1} +M_{\ a}^b \bar{r}_b r^a -\frac{i}{k} (\tilde{\mu}_{I+1\, \dot{1}}{}^{\dot{1}} - \tilde{\mu}_{I+1\, \dot{2}}{}^{\dot{2}}) - \frac{1}{2}
\end{pmatrix},
\end{equation}
with
\begin{equation}
\label{1/4M}
M=\Pi^{-1}\begin{pmatrix}
\frac{i}{k} & 0\\
0 & \bar{\beta}_2 \beta^{2} -\frac{i}{k}
\end{pmatrix}.
\end{equation}
In addition to the supercharge $Q$, these hyperloops preserve a supercharge $Q'$ arising from same 
$\eta^\imath_a$ but with $\bar\eta^\imath_a\to-\bar\eta^\imath_a$. The argument is identical 
to the case of the bosonic loops presented in Section~\ref{sec:bosonic}. 
In this case we see that the fermionic terms are unchanged if we 
keep the same $\beta$'s and $M\to-M$, so the diagonal entries $M_1^{\ 1} r^1 \bar{r}_1$ and 
$M_2^{\ 2} r^2 \bar{r}_2$ are also left invariant. 
The requirement that $M$ is diagonal guarantees, therefore, that the loop is also invariant 
under $Q'$ and is 1/8 BPS.

Thus, for any choice of $Q$ with $\Pi\neq0$, if we restrict the parameters such that $\beta^1=\bar\beta_1=0$, we 
find a family of 1/8 BPS hyperloops parametrized by $\beta^2$ and $\bar\beta_2$. However, as we can 
conjugate $\cL$ by a constant matrix
\beq
\cL\to \begin{pmatrix}1&0\\0&x^{-1}\end{pmatrix}\cL\begin{pmatrix}1&0\\0&x\end{pmatrix}\,,
\eeq
this gauge transformation eliminates one of the parameters, and we end up with a one (complex) dimensional 
moduli space.

This is very similar to the discussion in \cite{Drukker:2020dvr}, but it is much more general, as it works
with any of the supercharges $Q$ in \eqref{Q} with $\Pi\neq0$.
To make contact with the constructions in \cite{Drukker:2020dvr} we can look at the moduli space
of 1/4 BPS hyperloops studied there, which are all deformations of
the usual bosonic Gaiotto-Yin loops \cite{Gaiotto:2007qi}. Those
loops preserve a one-dimensional conformal group, under which the supercharges are charged. 
Looking at the algebra \eqref{QQ'} and requiring only conformal transformations in the square of the
supercharge imposes
$\epsilon_{\imath \jmath} \left( \bar\eta_a^\imath \eta_b^\jmath + \bar\eta_b^\imath \eta_a^\jmath \right) =0$.
To realize this, we choose two vectors $\bar w_a$ and $w_a$ (as usual, bar does not indicate complex conjugation).
For an arbitrary vector $s^\imath$, define parameters $\bar\eta, \eta$ as
\beq
\label{Q2Conformal}
\eta^\imath_a=w_a s^\imath \,,
\qquad
\bar\eta^\imath_a=\bar w_a s^\imath \,.
\eeq
The resulting supercharges are all linear combinations of
\beq
\label{1/4Q}
w_a Q_\imath^{\dot2 a%+
}\,,\qquad
%w_a Q_r^{\dot2 a\alpha}\,,\qquad
\bar w_a Q_{\bar \imath}^{\dot1 a %-
}\,,\qquad
%\bar w_aQ_{\bar l}^{\dot1 a\alpha}\,,
\eeq
whose anticommutators generate the bosonic algebra $\sof(2,1) \oplus \uni(1)$, where
the $\uni(1)$ summand is generated by $L_\perp + \frac{1}{2} w_a \bar w_b R^{ab}$ (see Section~\ref{sec:alg1} for details).

In \cite{Drukker:2020dvr} the vector $w_a$ was $\delta_a^2$ and $\bar w_a$ was $\delta_a^1$.
Other choices can be achieved by an $SU(2)_L$ rotation. What was more restrictive there is that
only a single choice of $Q$ (or $s^\imath$) was used. As long as we turn on only the parameters as in~\eqref{1/4M}, we preserve
all the supercharges in \eqref{1/4Q}, so any choice (with $\Pi\neq0$) is equivalent. When turning
on more $\beta$ parameters, we find different moduli spaces, depending on the exact choice
of $Q$. Our analysis here therefore generalizes also this simple case of deformations of the
1/4 BPS bosonic loop.

As discussed in Section~\ref{sec:bosonic}, there are several new bosonic loops generated 
by our construction that are not related to those in \cite{Drukker:2020dvr}. Clearly their deformations 
with $\beta\neq0$ are also new.

%%%%%%%%%%%%%%%%

\subsection{Two-node hyperloops with $\Pi=0$}
\label{sec:casesPi=0}

This case is presented in Section~\ref{sec:2-node-0}, where it is shown that 
the general deformation is of the form~\eqref{Q=0L} with 
$G$ and $C$ as in \eqref{Pi=0G}, subject to the constraints that $\bar\beta_\parallel=\beta^\parallel$ 
and the conditions on $\bar\beta_\perp$, $\beta^\perp$ and $c$ in \eqref{detT}. 
The resulting expression for $\cL$ is then in~\eqref{Pi=0loops}.

A simple way to find loops with enhanced supersymmetry is when the superconnection is invariant under 
$\su(2)_L$, which arises when $M_a^{\; b} r^a \bar{r}_b \propto \nu_I$.
Looking at the expression for $\nu$ in \eqref{Pi0nu} 
and $M$ in \eqref{Pi0M}, we see that one needs to impose
\beq
\label{su(2)Conditions}
\beta^\parallel=0\,,\qquad
\xi\alpha \bar{\beta}_{\perp} =\bar{\alpha} \beta^{\perp}\,.
\eeq
These equations are consistent with~\eqref{detT}%
\footnote{\eqref{detT} is also solved with $\xi=\xi_0e^{-i\varphi}$, with a constant $\xi_0\neq0$, 
arbitrary $\beta^\perp$, $\bar\beta_\perp$, $\beta^\parallel=\bar\beta_\parallel$ and $c=0$.}, 
combining all the parameters to a single periodic function $\gamma=1-ik\Lambda\bar\alpha\beta^\perp$ 
appearing in the superconnection as
\begin{equation}
\label{1/8loops}
\mathcal{L}= \begin{pmatrix}
A_{\varphi,I} +\frac{i}{k} \gamma \nu_I -\frac{i}{k} (\tilde{\mu}_{I\; \dot{1}}^{\; \dot{1}} - \tilde{\mu}_{I\; \dot{2}}^{\; \dot{2}}) & -\frac{i\bar{\alpha}}{2}(\gamma +1) \psi_{\dot{1}-} -\frac{i\bar{\alpha}}{2}(\gamma -1)\xi^{-1} \psi_{\dot{2}+}\\
\frac{i \alpha}{2} (\gamma +1) \bar{\psi}_+^{\dot{1}} -\frac{i \alpha}{2} (\gamma -1)\xi \bar{\psi}_-^{\dot{2}}& 
A_{\varphi,I+1} +\frac{i}{k} \gamma \nu_{I+1} -\frac{i}{k} (\tilde{\mu}_{I+1\dot{1}}^{\quad\;\; \dot{1}} -\tilde{\mu}_{I+1\dot{2}}^{\quad\;\; \dot{2}})+c -\frac{1}{2}
\end{pmatrix},
\end{equation}
and $c=i\frac{\gamma -1}{2}\partial_{\varphi}\log (\xi e^{i\varphi})$. 

The degree of supersymmetry enhancement depends on the choice of supercharge $Q$. Specifically, 
following Section~\ref{sec:alg2}, we distinguish three cases.

%%%%%%%%%%%%

\subsubsection{1/8 BPS loops}
\label{sec:marcia1/8}

First, suppose $0 \neq Q^2 \in \su(2)_L$. Putting together (\ref{etatw-etatz}) and (\ref{etasw}), one sees that 
the parameters $\eta$, $\bar\eta$ may be cast into the form 
\begin{align}
\eta^\imath_a = t^\imath \, w_a\,, 
\qquad 
\bar\eta^\imath_a = \bar t^\imath \, w_a\,,
\end{align}
with some vector $w_a \neq 0$ and $\epsilon_{\imath\jmath} t^\imath \bar t^\jmath \neq 0$. 
Acting on the resulting supercharge with $\su(2)_L$, we find that, regardless of the choice of 
$w_a$, the loop preserves the two supercharges (with a convenient normalization)
\bal 
Q_1 =\frac{1}{\sqrt{\epsilon_{\imath\jmath} t^\imath \bar t^\jmath}}
\left( t^\imath \, Q_\imath^{{\dot 2}1} + \bar t^\imath \,(\sigma_1)\indices{_\imath^{\bar\jmath}}\, Q^{{\dot 1}1}_{\bar \jmath}\right),
\qquad
Q_2 = \frac{1}{\sqrt{\epsilon_{\imath\jmath} t^\imath \bar t^\jmath}}
\left(t^\imath \, Q_\imath^{{\dot 2}2} + \bar t^\imath \,(\sigma_1)\indices{_\imath^{\bar\jmath}}\, Q^{{\dot 1}2}_{\bar \jmath}\right).
\eal
Using~\eqref{anticomms} it is easy to verify that their anticommutators generate $\su(2)_L$
\bal
\{Q_1,Q_1\} = \frac{1}{2}  R_+\,, \qquad
\{Q_1,Q_2\} = - R_3\,, \qquad
\{Q_2,Q_2\} = - \frac{1}{2}  R_-\,.
\eal

%%%%%%%%%%%

\subsubsection{1/4 BPS loops and conformal loops}
\label{sec:marcia1/4}

Another case is when the supercharge satisfies $0 \neq Q^2 \in \mathfrak{u}(1)_{L_\perp}$. 
In this case, as derived in~\eqref{eqn:Q2U(1)xi}, we have $\xi = \xi_0 e^{-i\varphi}$, which immediately 
implies $c = 0$. As discussed in Section~\ref{sec:alg2}, the parameters of $Q$ take the form 
\bal
\label{eta1/4}
\eta^l_a &= t s^l w_a\,,
&\qquad
\eta^r_a &= t s^r z_a\,,
\\
\bar\eta^l_a &= \bar t s^l w_a\,,
&\qquad
\bar\eta^r_a &= \bar t s^r z_a\,,
\eal
where both $\epsilon^{ab}w_a z_b \neq 0$ and $s^l s^r \neq 0$. 
Knowing that the loop is invariant under $\su(2)_L$ we can project \eqref{eta1/4} to those terms 
involving either only $w_a$ or only $z_a$. Acting then with raising and lowering operators projects 
further to the two components $a=1, 2$, removing the dependence on $w_a$ and $z_a$ altogether, 
and leaving us with four supercharges
\bal
\label{14BPS}
Q_1 &= t Q^{{\dot 2} 1}_r + \bar t Q^{{\dot 1}1}_{\bar l} \,, &\qquad
Q_2 &= t Q^{{\dot 2} 1}_l + \bar t Q^{{\dot 1}1}_{\bar r} \,, 
\\
Q_3 &= t Q^{{\dot 2} 2}_r + \bar t Q^{{\dot 1}2}_{\bar l} \,, 
&\qquad
Q_4 &= t Q^{{\dot 2} 2}_l + \bar tQ^{{\dot 1}2}_{\bar r} \,.
\eal
Examining these, we see that they form doublets of $\sof(2,1)$ (exchanging $l$ and $r$).

The algebra generated by these supercharges is very simple, with the only non-vanishing anticommutators
\bea
\label{smallalgebra}
\{Q_1, Q_4\} = -2t\bar tL_\perp\,,
\qquad
\{Q_2, Q_3\} = 2 t \bar t L_\perp\,.
\eea
Note that the bosonic part of this $1/4$ BPS algebra is just $\mathfrak{u}(1)_{L_\perp}$, while $\su(2)_L$ and 
the one-dimensional conformal algebra $\sof(2,1)$ act as outer automorphisms. 

We noted that the superconnection \eqref{1/8loops} is invariant under $\su(2)_L$. It is interesting to check 
whether it is also invariant under $\sof(2,1)$. This clearly requires $\gamma$ to be a constant, as otherwise 
$\cL$ is not invariant even under rotations. Considering then a general conformal generator 
$J=a_+ J_++a_0 J_0+a_- J_-$ and using (\ref{J0+-Bosons})-(\ref{J0+-Fermions}), one finds that 
the conformal transformation of $\cL$ in (\ref{1/8loops}) is a total derivative
\bea
\label{ConfInv}
J {\cal L} = {\cal D}_\varphi^{\cal L}(a {\cal L}+H),
\eea 
with 
\bea
a=a_+e^{i\varphi}-ia_0+a_- e^{-i\varphi}, \qquad H=\left(\begin{array}{cc}0 &0 \\ 0 & a/2\end{array}\right).
\eea
The resulting Wilson loops are then invariant under 
$\sof(2,1) \oplus \su(2)_L \oplus \mathfrak{u}(1)_{L_\perp}$, providing a previously unidentified family 
of conformal 1/4 BPS loops. 

Note that the argument here is classical and as the superalgebra \eqref{smallalgebra} does not include the 
conformal generators, we cannot be sure that it is not spoiled by quantum corrections.

%%%%%%%%%%%

\subsubsection{Further 1/8 BPS loops}

The last example arising from \eqref{1/8loops} are loops with nilpotent $Q$. Since this case lies at the intersection 
of the previous two, we have to impose all the conditions discussed above. For the parameters, we have
\bal
\eta^\imath_a = a\rho^\imath w_a\,,
\qquad
\bar\eta^\imath_a = \bar a \rho^\imath w_a\,.
\eal
They give a pair of nilpotent supercharges 
\bal
Q_1 &= a \rho^\imath Q^{\dot{2}1}_\imath + \bar a \rho^\imath (\sigma_1)\indices{_\imath^{\bar \jmath}} Q^{\dot{2}1}_{\bar\jmath}\,, 
\\
Q_2 &= a \rho^\imath Q^{\dot{2}2}_\imath + \bar a \rho^\imath (\sigma_1)\indices{_\imath^{\bar \jmath}} Q^{\dot{2}2}_{\bar\jmath}\,, 
\eal
whose anticommutator vanishes as well.

Another family of loops with enhanced supersymmetry arises if, instead of $\su(2)_L$ symmetry 
(as in \eqref{1/8loops}), we demand conformal invariance from the beginning. Generalising the discussion 
in Section~\ref{sec:marcia1/4}, we impose the equation~\eqref{ConfInv} directly on the superconnection~\eqref{Pi=0loops}. 
The off-diagonal components of this matrix equation are satisfied, as in Section~\ref{sec:marcia1/4}, 
as long as $\xi=\xi_0 e^{-i\varphi}$ and
$c=0$, which identically solves the supersymmetry conditions~\eqref{detT}. Additionally, if we redefine
\begin{align}
\beta^\perp = \frac{\alpha}{2\Lambda} (\gamma-1), \qquad \bar\beta_\perp = \frac{\bar\alpha}{2\Lambda \xi}(\bar\gamma - 1), \qquad 
\beta^\parallel = \frac{i}{k\Lambda} \gamma^\parallel,
\end{align}
then we need to impose that $\gamma$, $\bar\gamma$ and $\gamma^\parallel$ are constants. 

The expression for the superconnection~\eqref{Pi=0loops} then becomes
\begin{equation}
\label{conformal}
\mathcal{L}= \begin{pmatrix}
A_{\varphi,I} + M_a{}^b r^a \bar{r}_b -\frac{i}{k} (\tilde{\mu}_{I\; \dot{1}}^{\; \dot{1}} - \tilde{\mu}_{I\; \dot{2}}^{\; \dot{2}}) & -\frac{i\bar{\alpha}}{2}(\bar\gamma +1) \psi_{\dot{1}-} -\frac{i\bar{\alpha}}{2}(\bar\gamma -1)\xi^{-1} \psi_{\dot{2}+}\\
\frac{i \alpha}{2} (\gamma +1) \bar{\psi}_+^{\dot{1}} -\frac{i \alpha}{2} (\gamma -1)\xi \bar{\psi}_-^{\dot{2}}& 
A_{\varphi,I+1} +M_a{}^b \bar{r}_b r^a-\frac{i}{k} (\tilde{\mu}_{I+1\dot{1}}^{\quad\;\; \dot{1}} -\tilde{\mu}_{I+1\dot{2}}^{\quad\;\; \dot{2}}) -\frac{1}{2}
\end{pmatrix},
\end{equation}
with the couplings to the rotated scalars~\eqref{Pi0M} given by
\begin{equation}
M_a{}^b= \frac{i}{k\Lambda}\begin{pmatrix}
0 & \bar\gamma\\
\gamma & \gamma^{\parallel}
\end{pmatrix}.
\end{equation}
The remaining check is whether the diagonal part of equation~\eqref{ConfInv} is satisfied, which imposes that 
the couplings to the unrotated scalars $q^a, \bar q_a$ are constant. 
This can be arranged in two ways. Firstly, by~\eqref{Pi0nu} we can set $\bar\gamma = \gamma, \gamma^\parallel=0$ to obtain 
scalar terms proportional to $\nu_I, \nu_{I+1}$ without any explicit $\varphi$ dependence. These loops are 
just the conformal $1/4$ BPS loops described in the previous section. 

Alternatively, constant scalar couplings can be 
obtained for arbitrary $\gamma, \bar\gamma, \gamma^\parallel$ by demanding instead 
$\epsilon^{ab}\bar\eta^l_a \bar\eta^r_b=0$ or, equivalently, $Q^2 = 0$. 
In order to derive the symmetries preserved by these loops, we parametrise the supercharge using~\eqref{nilpotent} 
and act on it with the conformal generators. This process generates another supercharge, so in total we have
\begin{align}
Q_1 = w_a\left( a Q^{{\dot 2}a}_l + \bar a Q^{{\dot 1}a}_{\bar r} \right), 
\qquad 
Q_2 = w_a\left( a Q^{{\dot 2}a}_r + \bar a Q^{{\dot 1}a}_{\bar l} \right).
\end{align}
Both these supercharges are nilpotent and their anticommutator vanishes. By construction, $\sof(2,1)$ acts on the algebra 
as an outer automorphism.

There is yet another example of supersymmetry enhancement without $\su(2)_L$ symmetry, but with invariance under 
$T_\perp$ (but not $L_\perp$ in \eqref{Lperp}). Recalling that $T_\perp$ acts diagonally and
separates barred from unbarred supercharges, it is easily seen that the commutator $Q'=[T_\perp, Q]$ 
is linearly independent of $Q$, provided $Q$ comprises both barred and unbarred supercharges 
(so $\xi\neq0,\infty$). 
To see which loops are invariant under $Q'$, we note that keeping $\eta^\imath_a$ and changing 
$\bar\eta^\imath_a\to-\bar\eta^\imath_a$ leaves $\Pi=0$, likewise $\Lambda$ is unmodified, and $\xi\to-\xi$. 
Noticing that \eqref{Pi=0loops} contains terms proportional to both $\Lambda\beta^\perp$ and 
$\xi\Lambda\beta^\perp$, we have to set $\beta^\perp=0$ and similarly for $\bar \beta_\perp$, which by \eqref{detT} also fixes $c=0$.
The resulting superconnection is
\begin{equation}
\label{Pi=0loopscase1}
\mathcal{L}=\cL_{1/2}
+ \begin{pmatrix}
\beta^{\parallel} r^{\parallel} \bar{r}_{\parallel}&0\\
0&\beta^{\parallel} \bar{r}_{\parallel}r^{\parallel} 
\end{pmatrix},
\end{equation}
where $\beta^\parallel$ can be an arbitrary periodic function of $\varphi$. One can check that generically 
the supersymmetry is not enhanced further.

%%%%%%%%%%%

\subsubsection{The special cases: $\xi=0$ and $\xi=\infty$}

When $\xi=0$, the superconnection of loops are the same as \eqref{Pi=0loops} with $\xi=c=0$ and 
$\beta^{\perp}$, $\bar{\beta}_{\perp}$ and $\beta^\parallel$ free. If we want to study the $\su(2)_L$ 
enhanced points, we should impose $\beta^{\perp} =\beta^{\parallel}=0$ and get the loops
\begin{equation}
\label{t=0loops}
\mathcal{L}=\mathcal{L}_{1/2}+\begin{pmatrix}
0 & -i\Lambda \bar{\beta}_{\perp} \psi_{\dot{2}+}\\
0 & 0
\end{pmatrix}.
\end{equation}
The case $\xi=\infty$ is similar with a term on the lower left corner.

In all of these examples the free parameters $\beta^\parallel$, $\beta^\perp$ (and in the last case also 
$\bar\beta_\perp$) are any periodic functions of $\varphi$. The reason is most transparent with regards 
to $\beta^\parallel$, as $Q$ annihilates $r^\parallel\bar r_\parallel$ and we can insert any 
density of them along the loop.

In Sections~\ref{sec:bosonic}, \ref{sec:cases2nodes} and \ref{sec:casesPi=0} above, we noted multiple examples 
of hyperloops that in addition to $Q$ preserve also $Q'$ with $\bar\eta^\imath_a\to-\bar\eta^\imath_a$. 
They clearly also preserve $Q\pm Q'$, which are supercharges with $\Pi=0$ and $\xi=0$ and $\xi=\infty$. 

%%%%%%%%%%%%%%%%%%%

\subsection{Hyperloops with twisted hypers and $\Pi\neq0$}
\label{sec:cases-twisted}

To couple our hyperloops to the twisted hypermultiplets, the starting point in Section~\ref{sec:longer} 
is a $4\times4$ superconnection \eqref{4nodeL1/2} which takes a block-diagonal form and is 
deformed with parameters $\beta$ and $\delta$. Here we focus on 
special examples of these loops. As a first step, we set all the $\beta$'s to zero. In the absence of 
the $\delta$ terms, this would give a diagonal connection with only bosonic fields.

With $\beta=0$ and $\delta\neq0$, we find instead a block-diagonal form, with a $2\times2$ block 
involving the nodes $I+1$ and $I+2$, and two decoupled nodes $I$ and $I+3$. We ignore in the following the 
decoupled nodes and concentrate only on the remaining $2\times2$ block. Note that often the decoupled 
nodes do not preserve the symmetries of the central block. This can be remedied in the setting of a 
circular quiver.

In the case of a deformation with $\delta_{I+1\,\dot 1}$ and $\bar\delta^{\dot1}_{I+1}$, the central block takes the form
\beq
\cL=\begin{pmatrix}
A_{\varphi, I+1} +M^b_{\ a} \bar{r}_{I\,b} r^a_I + \widetilde{M}^{\dot a}_{\ \dot{b}} \tilde{q}_{I+1\,\dot{a}} \bar{\tilde q}_{I+1}^{\dot b} - \frac{1}{2} &-i\bar{\delta}^{\dot 1}_{I+1}\tilde{\rho}_{I+1\,+}^2
\\
-i \Pi^{-1}\delta_{I+1\,\dot 1} \bar{\tilde\rho}_{I+1\,2-} 
&A_{\varphi,I+2} + M_a^{\ b} r_{I+2}^a \bar{r}_{I+2\,b}
+ \widetilde{M}^{\ \dot a}_{\dot{b}} \bar{\tilde q}^{\dot b}_{I+1} \tilde{q}_{I+1\,\dot{a}} + \bar\Gamma 
\end{pmatrix}
\eeq
with (see \eqref{Mrr})
\beq
M=\frac{i}{k}\Pi^{-1}\begin{pmatrix}1&0\\0&-1\end{pmatrix}\,,
\qquad
\widetilde{M}= \begin{pmatrix}
-i/k + {\bar\delta}^{\dot 1}_{I+1}\delta_{I+1\,\dot 1} & 0 \\
0 & i/k
\end{pmatrix}.
\eeq
This structure is the analog of the two-node quiver with a coupling to a single pair of scalars 
in the hypermultiplets as in \eqref{1/4M}. Just as in that example, these loops have enhanced 
supersymmetry with the second supercharge $Q'$ given by exchanging 
$\bar\eta^\imath_a\to-\bar\eta^\imath_a$. So all these loops are at least $1/8$ BPS.

Further supersymmetry enhancement arises in the schemes explained in Section~\ref{sec:bosonic} leading to operators that can preserve either 3 or 4 supercharges. Even further supersymmetry enhancement arises by setting 
$\delta_{I+1\,\dot{1}}\bar\delta_{I+1}^{\dot 1}=2i/k$, as then the loop enjoys $\su(2)_R$ symmetry. 
In this case, the analysis of the previous paragraph is extended to supercharges with $\dot1\leftrightarrow\dot2$
and we have a doubling of the amount of preserved supersymmetry. 
Note that because of the $\dot1\leftrightarrow\dot2$ exchange, these supercharges are not 
preserved by the original 1/2 BPS loop, see \eqref{1/2susy}. The $1/4$ BPS loop becomes the $1/2$ BPS operator coupling to the single pair of scalars $\tilde{q}_{\dot 1}$, $\bar{\tilde{q}}^{\dot 1}$ from the twisted hypermultiplet. The $1/8$ BPS operator becomes $1/4$ BPS and for the particular parameterization
\beq
\label{latitudeparameters} 
\bar\eta_1^r=\eta_2^l=\cos\frac{\theta}{2}\,,
\qquad
\bar\eta_2^l=-\eta_1^r=\sin\frac{\theta}{2}\,,
\eeq
we recover the ``fermionic latitude'' loops constructed first in ABJM theory in \cite{Cardinali:2012ru} and
generalized to $\cN=4$ theories in \cite{Drukker:2020dvr}, see also \cite{Bianchi:2018bke}. The $3/16$ BPS operator becomes $3/8$ BPS.

Completely analog constructions arise with $\delta_{I+1\,\dot1}=\bar\delta^{\dot1}_{I+1}=0$ and 
nonzero couplings $\delta_{I+1\,\dot2}$ and $\bar\delta^{\dot2}_{I+1}$. The most symmetric loop of this 
class is the second 1/2 BPS loop coupling instead to the pair of scalars $\tilde{q}_{\dot 2}$, $\bar{\tilde{q}}^{\dot 2}$.
The cases with all four $\delta$ parameters non-vanishing is allowed, as long as \eqref{obstruction} is satisfied.
The analysis follows as before, but $\su(2)_R$ symmetry is preserved only when we restrict to a single pair of $\delta$. 

%%%%%%%%%%%%%%%%%%%%%

\subsection{Hyperloops with twisted hypers and $\Pi=0$}
\label{sec:cases-twistedPi0}

These operators are considered in Section~\ref{sec:4node-0}, where we find supersymmetric loops built out of the $G$ and $C$ in \eqref{4nodeGC-0}. In particular, the $\beta$ parameters that couple to scalars from the untwisted hypermultiplet satisfy the same constraints as in the 2-node case, while the couplings to the twisted scalars, $\bar{d}^{\dot a}$ and $d_{\dot a}$, are arbitrary periodic functions as long as $\lambda=0$.

Denoting the superconnection in \eqref{Pi=0loops} as $\cL_{\Pi=0}$, the expression we find for $\cL$ is
\beq
\label{4nodesPi=0loops}
\cL = \begin{pmatrix}
\cL_{\Pi=0} & \begin{matrix}
\bar\alpha_I \bar{d}^{\dot a}r_I^\parallel \tilde{q}_{I+1\,\dot a} & 0 \\
\bar{d}^{\dot 1}\tilde\rho_{I+1,+}^2 + \bar{d}^{\dot 2}\tilde\rho_{I+1,-}^1 & \bar{d}^{\dot a}\bar\alpha_{I+2} \tilde{q}_{I+1\,\dot a}r_{I+2}^\parallel
\end{matrix} \\
\begin{matrix}
-d_{\dot a}\alpha_I \xi \bar{\tilde{q}}^{\dot a}_{I+1} \bar{r}_{I\parallel} & d_{\dot 1}\bar{\tilde{\rho}}_{I+1,2-}+d_{\dot 2}\bar{\tilde{\rho}}_{I+1,1+} \\
0 & -\alpha_{I+2} d_{\dot a} \xi \bar{r}_{I+2\parallel}\bar{\tilde{q}}^{\dot a}_{I+1}
\end{matrix} & \cL_{\Pi=0}
\end{pmatrix}.
\eeq
Note that the coupling to the twisted scalar bilinears is unchanged and the $\widetilde{M}$ in the central nodes does not receive contributions from the $d$'s. In general, these loops preserve a single supercharge. 

One special case is similar to the $1/4$ BPS hyperloop of Section~\ref{sec:casesPi=0}, when $\xi$ \eqref{xipar} is of 
the form $\xi_0 e^{-i\varphi}$ with constant $\xi_0$. This can arise with either $\xi_0=\eta^r/\bar\eta^r$ or 
$\xi_0=\eta^l/\bar\eta^l$ leading to a two fold degeneracy. This is a symmetry of the 
superconnection \eqref{4nodesPi=0loops} when
\beq
\bar{d}^{\dot 1}=d_{\dot 2}=\frac{1}{(\eta\bar{v})_1}\,,\quad \bar{d}^{\dot 2}= d_{\dot 1}= \frac{1}{(\bar\eta v)_1}\,,
\eeq
and $(\bar\eta v)_1=(\bar\eta v)_2$, $(\eta \bar{v})_1=(\eta \bar{v})_2$. The resulting hyperloop preserves 
two supercharges and, as before, $\sof(2,1)$ acts as an outer automorphism on the preserved superalgebra. 
Unlike the 2-nodes case in \eqref{1/8loops}, there is no way to restore $\su(2)_L$ symmetry and find 
further supersymmetry enhancement.

%%%%%%%%%%%

\subsubsection{The special cases: $\xi=0$ and $\xi=\infty$}
\label{sec:cases-twistedxi0}

In Section~\ref{sec:4-node-xi0} the analysis of the case of $\xi=0$ is extended to include the twisted hypermultiplets. Denoting the superconnection in \eqref{xi=0} as $\cL_{\xi=0}$, the extension to include twisted hypermultiplets gives
\beq
\cL = \begin{pmatrix}
\cL_{\xi=0} & \begin{matrix}
r_I^\parallel (\gamma_1 \tilde{q}_{I+1\,\dot{2}} + \bar\alpha_I \bar{d}^{\dot 1} \tilde{q}_{I+1\,\dot{1}}) & 0 \\
-i \bar{d}^{\dot 1} \tilde\rho^2_{I+1,+} & \frac{\bar\alpha_{I+2}}{\bar\alpha_I}(\gamma_1 \tilde{q}_{I+1\,\dot{2}} + \bar\alpha_I \bar{d}^{\dot 1} \tilde{q}_{I+1\,\dot{1}}) r_{I+2}^\parallel
\end{matrix} \\
\begin{matrix}
0 & (\delta_{I+1}-i d_{\dot 2})\bar{\tilde\rho}_{I+1,1+} \\
0 & 0
\end{matrix} & \cL_{\xi=0}
\end{pmatrix}\,.
\eeq
Note that, as $\delta_{I+1}$ and $d_{\dot 2}$ appear only through the linear combination $\delta_{I+1}-i d_{\dot 2}$, we can eliminate one of them. Supersymmetry enhancement relying on manifest $\su(2)_L$ symmetry happens only by setting to zero off-block-diagonal parameters, in which case we simply recover two decoupled copies of \eqref{t=0loops}.

%%%%%%%%%%%%%

\subsection*{Acknowledgements}

We would like to acknowledge fruitful discussions with L. Griguolo, E. Pomoni and Y. Wang. 
DT is grateful to the Galileo Galilei Institute for hospitality in the final stages of this work. 
ND is supported by the Science Technology \& Facilities Council under the grants  ST/T000759/1 and ST/P000258/1. 
The work of ZK is supported by CSC grant No. 201906340174.
MT acknowledges the support of the Conselho Nacional de Desenvolvimento Cientifico e Tecnologico (CNPq). 
DT is supported in part by the INFN grant {\it Gauge and String Theory (GAST)} and would like to thank 
FAPESP's partial support through the grants 2016/01343-7 and 2019/21281-4. 

%%%%%%%%%%%%%

\appendix

%%%%%%%%%%%%%

\section{Symmetries of the 1/2 BPS Wilson loop}
\label{app:SUSYtrans}

We start by recalling that the symmetries of an $\cN=4$ superconformal theory on $S^3$ form an $\osp(4|4) \cong D(2,2)$ superalgebra, with the bosonic symmetries $\sof(4,1) \oplus \sof(4).$ These are, respectively, the three-dimensional conformal algebra and the R-symmetry algebra. The latter is conveniently thought of as $\sof(4) \simeq \su(2)_L \oplus \su(2)_R$. The 16 supercharges transform as conformal spinors under $\sof(4,1)$ and in the fundamental representations of both R-symmetry $\su(2)$'s. 

The circular $1/2$ BPS loop breaks part of these symmetries. Specifically, of the conformal generators, it preserves only the one-dimensional conformal algebra along the contour of the loop and the rotations in the plane perpendicular to it
\bea
\sof(2,1)
\oplus \uni(1)_\perp\,.
\eea
$\su(2)_L$ is preserved by the loop, whereas $\su(2)_R$ is broken to $\uni(1)_R$.\footnote{Of course, the choice of which of the R-symmetry factors is broken and which one is preserved is a matter of which 1/2 BPS loop one considers, as explained in \cite{Cooke:2015ila}.} 

We denote the conformal generators along the circle by $J_0$ and $J_\pm$, with nonvanishing commutators 
\beq
[J_0, J_\pm] = \pm J_\pm\,,
\qquad
[J_+, J_-] = 2 J_0\,.
\eeq
Parametrising the circle by the angular coordinate $\varphi$, these generators can be represented by differential operators 
\bea
J_0 = - i \partial_\varphi,
\qquad
J_\pm = e^{\pm i \varphi} \partial_\varphi\,.
\eea
The action on the fields can be obtained by evaluating the usual conformal transformations on the circle. 
Suppressing R-symmetry indices, we find for the bosonic fields involved in our Wilson loops
\bal
\label{J0+-Bosons}
J_0 A_{\varphi} &= -i \partial_\varphi A_{\varphi}\,,
&\qquad
J_\pm A_{\varphi} &= e^{\pm i\varphi} \left(\partial_\varphi \pm i \right)A_{\varphi}\,,
\\
J_0 q &= -i\partial_\varphi q\,, 
&\qquad
J_\pm q &= e^{\pm i \varphi} (\partial_\varphi \pm i/2 )q\,, 
\\
J_0 \bar q &= -i \partial_\varphi \bar q\,,
&\qquad
J_\pm \bar q &= e^{\pm i \varphi}(\partial_\varphi \pm i/2 )\bar q\,.
\eal
The second term in the action of $J_\pm$ picks up the scaling dimension of the respective fields. Similarly, for the fermions
\bal
\label{J0+-Fermions}
J_0 \psi &= -i \left( \partial_\varphi + i \sigma_3 /2 \right) \psi\,, 
&\qquad
J_\pm \psi &= e^{\pm i \varphi} (\partial_\varphi \pm i + i \sigma_3/2)\psi\,, 
\\
J_0 \bar \psi &= -i \left( \partial_\varphi + i \sigma_3 /2 \right) \bar \psi \,,
&\qquad
J_\pm \bar \psi &= e^{\pm i \varphi}(\partial_\varphi \pm i + i \sigma_3/2 )\bar \psi\,.
\eal
We denote by $T_\perp$ the generator of rotations $\uni(1)_\perp$ in the orthogonal plane to the contour, which commutes with all other preserved conformal generators. The normalization of $T_\perp$ is fixed such that 
\bal
\label{Tperp}
T_\perp\psi &= \frac{i}{2}\sigma_3 \psi\,,
&\qquad
[T_\perp, \Q{a}{a}{\imath}] &= \frac{i}{2} \Q{a}{a}{\imath}\,,
\\
T_\perp\bar\psi &= \frac{i}{2}\sigma_3 \bar\psi\,,
& \qquad
[T_\perp, Q^{\dot{a} a}_{\bar \imath} ] &= - \frac{i}{2} Q^{\dot{a} a}_{\bar \imath}\,.
\eal

The generators of $\su(2)_L$ are $R_\pm, R_3$, with commutation relations
\begin{align}
[R_3, R_\pm] = \pm R_\pm\,,
\qquad
[R_+, R_-] = 2 R_3\,.
\end{align}
As mentioned above, these symmetries are preserved by the loop. We distinguish $\su(2)_R$ with bars: 
$\bar R_\pm, \bar R_3$. Only $\bar R_3$ is preserved by the loop. It is also useful to defined the twisted generator
\bea
\label{Lperp}
L_\perp \equiv -i \left( T_\perp +\frac{i}{2} \bar R_3 \right),
\eea
which mixes the rotations in the perpendicular plane in $\uni(1)_\perp$ with the $R$-symmetry rotations in $\uni(1)_R$ \cite{Agmon:2020pde}. 

The supercharges preserved by the loop are given in \eqref{1/2susy} and anticommute to
\beq
\begin{aligned}
\label{anticomms}
\{ \Q{2}{a}{l} , \barQ{1}{b}{l} \} &= \epsilon^{ab} \left( J_0 + L_\perp \right) + R^{ab}\,,&
\{ \Q{2}{a}{l} , \barQ{1}{b}{r} \} &= \epsilon^{ab} J_+\,,\\
\{ \Q{2}{a}{r} , \barQ{1}{b}{l} \} &= - \epsilon^{ab} J_-\,,\\
\{ \Q{2}{a}{r} , \barQ{1}{b}{r} \} &= \epsilon^{ab} \left( J_0 - L_\perp \right) - R^{ab}\,.
\end{aligned}
\eeq
Here, we have contracted the $\su(2)_L$ generators with the Pauli matrices in the usual fashion and raised one index by $\epsilon^{ab}$ (with $\epsilon^{12}=1$), such that 
\bea
R^{ab} = 
\begin{pmatrix}
R_+ & - R_3 \\
 - R_3 & - R_-
\end{pmatrix}.
\eea
In order to fully specify the superalgebra, one computes the commutators of bosonic and fermionic generators using the super-Jacobi identities. Explicitly, we find 
that the residual conformal generators act on the supercharges as follows
\bal
\label{Jpm0Q}
J_+ \begin{pmatrix}
Q_l \\ Q_r
\end{pmatrix} &= \begin{pmatrix}
0 \\ -Q_l
\end{pmatrix},
&\qquad
J_+ \begin{pmatrix}
Q_{\bar l} \\ Q_{\bar r}
\end{pmatrix} &= \begin{pmatrix}
-Q_{\bar r} \\ 0
\end{pmatrix},
\\
J_- \begin{pmatrix}
Q_l \\ Q_r
\end{pmatrix} &= \begin{pmatrix}
-Q_r \\ 0
\end{pmatrix},
&\qquad
J_- \begin{pmatrix}
Q_{\bar l} \\ Q_{\bar r}
\end{pmatrix} &= \begin{pmatrix}
0 \\ -Q_{\bar l}
\end{pmatrix}, 
\\
J_0 \begin{pmatrix}
Q_l \\ Q_r
\end{pmatrix} &= \frac{1}{2} \begin{pmatrix}
Q_l \\ -Q_r
\end{pmatrix},
&\qquad
J_0 \begin{pmatrix}
Q_{\bar l} \\ Q_{\bar r}
\end{pmatrix} &= \frac{1}{2} \begin{pmatrix}
-Q_{\bar l} \\ Q_{\bar r}
\end{pmatrix}, 
\\
T_\perp \begin{pmatrix}
Q_l \\ Q_r
\end{pmatrix} &= \begin{pmatrix}
Q_l \\ Q_r
\end{pmatrix}, 
&\qquad
T_\perp \begin{pmatrix}
Q_{\bar l} \\ Q_{\bar r}
\end{pmatrix} &= -\begin{pmatrix}
Q_{\bar l} \\ Q_{\bar r}
\end{pmatrix}.
\eal
These (anti-)commutators together with the bosonic structure outlined above define the Lie superalgebra $\mathfrak{sl}(2|2)$.
As is easily checked, $L_\perp$ commutes with all supercharges as well as all bosonic generators. Indeed, $\mathfrak{sl}(2|2)$ is 
a central extension of the classical Lie superalgebra $A(1,1)$ by $\uni(1)$, so this structure is expected~\cite{Frappat:1996pb}.

%%%%%%%%%%%%%%

\section{The covariant derivative}
\label{app:SusyCondition}

Here we explain what it concretely means when a supersymmetry transformation on a superconnection ${\cal L}$ acts as a total covariant derivative, as in \eqref{QL1/2}
\beq
Q \cL = \cD^\cL_\varphi H\,.
\eeq

Consider the open Wilson loop (we shall worry about taking the supertrace later)
\beq
W_{2\pi,0} = \cP \exp i \int_0^{2\pi} d\varphi \,\cL \,,
\eeq
and act with $Q$ on the loop. 
It is crucial that the superconnection $\cL = \cL^B + \cL^F$ is an even supermatrix, 
{\it i.e.} a matrix whose diagonal entries $\cL^B$ are exclusively bosonic and whose off-diagonal entries $\cL^F$ are exclusively fermionic, and likewise for the Wilson loop. 
Commuting $Q$ through a product of two such superconnections ${\cal L}_1$ and ${\cal L}_2$, one gets 
$Q (\cL_2 \cL_1) = Q(\cL_2) \cL_1 + \sigma_3 \cL_2\sigma_3 Q\cL_1$, 
where the Pauli matrix is introduced to flip the sign of the odd part of $\cL_1$. 

Acting with $Q$ on $W_{2\pi,0}$, one needs to apply the Leibniz rule, as $Q$ can act on any $\cL(\varphi)$. 
Keeping track of the sign changes, one finds
\beq
QW_{2\pi,0}= i \sigma_3 \int_0^{2\pi} d\varphi\, W_{2\pi, \varphi} \left(\sigma_3 Q \cL(\varphi) \right)W_{\varphi, 0}\,.
%= i \sigma_3 W\left[\int_0^{2\pi} d\varphi\, \sigma_3 Q \cL(\varphi)\right].
\eeq
Now let us assume it exists an $H(\varphi)$, such that $Q\cL=\sigma_3\cD_\varphi^\cL(\sigma_3 H(\varphi))$. Then, by the standard relations for 
Wilson loops, one finds
\beq
QW_{2\pi,0}= i \sigma_3 \int_0^{2\pi} d\varphi\, W_{2\pi, \varphi} \cD_\varphi^\cL(\sigma_3 H(\varphi)) W_{\varphi, 0}
= i H(2\pi)W_{2\pi, 0}-i\sigma_3W_{2\pi,0}\sigma_3 H(0)\,.
\eeq
Assuming $H(\varphi)$ to be periodic and taking the supertrace, one gets
\beq
QW=i\sTr(H(0)W_{2\pi, 0}-\sigma_3W_{2\pi,0}\sigma_3 H(0))
=i\Tr([\sigma_3H(0),W_{2\pi, 0}])=0\,.
\eeq
This implies that the covariant derivative that should appear in the supersymmetry transformations is
\beq
\label{covariantderivative}
Q\cL=\sigma_3\cD_\varphi^\cL(\sigma_3 H)
=\partial_\varphi H-i[\cL_\textrm{bos},H]+i\{\cL_\textrm{fer},H\}\,.
\eeq
In the main text we write this as $\cD_\varphi^\cL H$, but we really mean the expression above with 
the anticommutator of the fermionic part of the superconnection.

If one prefers working instead with bosonic variations, one can introduce a Grassmann parameter $\xi$ and write $\delta = \xi Q$. 
The analogous supersymmetry condition reads
\beq
\delta \cL = \cD^{\cL}_\varphi (\xi H)\,.
\eeq

%%%%%%%%%%%%%%%%%%%%%%%%

\section{Extra fermionic terms}
\label{app:modification}

In this appendix we examine the possibility to add extra fermionic terms to the $F$ in the superconnection, beyond 
the term $-iQG$ in \eqref{Q=0L}. This term arises in the case of $\Pi=0$ in Section~\ref{sec:2-node-0}, where 
$G$ includes only two scalar fields \eqref{Pi=0G} and, consequently, $QG$ has only two linear combinations of the fermions 
\eqref{Pi0Qr}. To generalize it, we take an extra term related to the 
fermions in the original 1/2 BPS connection
\begin{equation}
\label{modified-F}
F=-iQ G + (D-1) \mathcal{L}_{1/2}^F\,.
\end{equation}
Here $D=\diag (\bar d,d)$.

The result of the analysis below is that such addition is only possible for $\xi=0$ or $\xi=\infty$, and 
those cases are already treated in Section~\ref{sec:2-node-xi}. So this appendix leads to no further 
hyperloops beyond those described in the main text.

Taking \eqref{modified-F} and using the same equations for the variations $QB$ and $QF$ in \eqref{BF-var}, 
one gets $\Delta H =(D-1) H$, because there is no derivative term in $Q^2 G$. Plugging everything known 
into \eqref{BF-var} yields
\begin{equation}
\begin{aligned}
-iQ^2 G &=(\partial_{\varphi} D)H-i[B+C, DH]\,,\\
QB &=\{QG,DH\} +i(\det D-1)\{\mathcal{L}_{1/2}^F,H\}\,.
\end{aligned}
\end{equation}
Focusing on the second equation for now and using $QH=0$ and 
$Q\mathcal{L}_{1/2}^B =i \{\mathcal{L}_{1/2}^F,H\}$, one gets
\begin{equation}
Q B=Q\{DH,G\} +(\det D-1) Q \mathcal{L}_{1/2}^B\,,
\end{equation}
which is simply solved by
\begin{equation}
\label{B}
B= \{DH,G\} +(\det D-1) \mathcal{L}_{1/2}^B\,.
\end{equation}
Extra terms annihilated by $Q$ are included in $C$.

The case of $\det D=1$ is simply a gauge transformation, changing $\alpha$ and $\bar\alpha$. So we are left 
with examining the case $\det D\neq1$. This results in $B$ having a term proportional to $\cL_{1/2}^B$, 
which includes the gauge fields. Since the gauge fields cannot appear in a Wilson loop with an arbitrary 
prefactor (they should have prefactor $i$), one needs to cancel part of this term with factors of the gauge field in $C$. This amounts to finding 
a connection annihilated by $Q$, which one can assume to be purely bosonic: 
$\cL^{B\prime}=\diag(\mathcal{A}_{I}', \mathcal{A}_{I+1}'-1/2)$. We take
\begin{equation}
\mathcal{A}_{I}' =A_{\varphi} -\frac{i}{k} (M_{a}{}^{b} r^{a} \bar{r}_{b} +\tilde{\mu}_{\dot{1}}^{\dot{1}} -\tilde{\mu}_{\dot{2}}^{\dot{2}})\,,
\end{equation}
where $a,b\in \{\parallel,\perp\}$, and the task is now to find the coefficient matrix $M_a{}^b$. Using
\bal
r^{\parallel} (\bar{\psi}_+^{\dot{1}} &+\xi\bar{\psi}_-^{\dot{2}}) 
-(\xi \psi_{\dot{1}-} -\psi_{\dot{2}+}) \bar{r}_{\parallel} 
\\&=\Lambda (M_{\parallel}{}^{\perp} r^{\parallel} +M_{\perp}{}^{\perp} r^{\perp}) (\bar{\psi}_+^{\dot{1}} -\xi\bar{\psi}_-^{\dot{2}}) 
-\Lambda (\xi \psi_{\dot{1}-} +\psi_{\dot{2}+}) 
(M_{\perp}{}^{\parallel} \bar{r}_{\parallel} +M_{\perp}{}^{\perp} \bar{r}_{\perp})\,,
\eal
and imposing $Q \mathcal{A}_{I}'=0$ results in
\begin{equation}
\begin{aligned}
%&Q \left(A_{\varphi} -\frac{i}{k}(M_{a}{}^{b} r^{a} \bar{r}_{b} +\tilde{\mu}_{\dot{1}}^{\dot{1}} -\tilde{\mu}_{\dot{2}}^{\dot{2}})\right)=0\\&
Q \left(A_{\varphi} -\frac{i}{k}(-\nu_I +\tilde{\mu}_{\dot{1}}^{\dot{1}} -\tilde{\mu}_{\dot{2}}^{\dot{2}})\right)=-\frac{2i}{k}(\xi \psi_{\dot{1}-} \bar{r}_{\parallel} +r^{\parallel} \bar{\psi}_+^{\dot{1}})\,,
\end{aligned}
\end{equation}
which is solved by
\begin{equation}
\label{t=0}
\xi=M_{\perp}{}^{\perp}=0,\quad M_{\parallel}{}^{\perp} =-M_{\perp}{}^{\parallel} =1/\Lambda\,.
\end{equation}
We see that indeed this works only for $\xi=0$ and therefore it falls under the cases already analyzed in 
Section~\ref{sec:2-node-xi}.

To compare with the analysis in Section~\ref{sec:2-node-xi}, 
we note that for $\xi=0$ there are many specific features, such as $Q^2 G=H^2=0$. 
We can also check that $\mathcal{L}_{1/2}^B -\mathcal{L}^B$ commutes with $DH$ and the only remaining 
supersymmetry conditions is
\begin{equation}
\partial_{\varphi} \bar d =-ic \bar d\,.
\end{equation}
Including the bosonic loop
\begin{equation}
\mathcal{A}_{I}'=A_{\varphi} -\frac{i}{k\Lambda} (\Lambda M_{\parallel}{}^{\parallel} r^{\parallel} \bar{r}_{\parallel} +r^{\parallel} \bar{r}_{\perp} -\bar{r}^{\perp} \bar{r}_{\parallel}) -\frac{i}{k} (\tilde{\mu}_{\dot{1}}^{\dot{1}} -\tilde{\mu}_{\dot{2}}^{\dot{2}})\,,
\end{equation}
and the analogous expression for $\cA_{I+1}'$ in $C$ with prefactors $1-\det D$ and combining all the terms, one finally gets the superconnection
\begin{equation}
\label{t=0L}
\mathcal{L}=\begin{pmatrix}
A_{\varphi,I} +M_a{}^{b} r^a \bar{r}_b -\frac{i}{k} (\tilde{\mu}_{I\; \dot{1}}^{\; \dot{1}} - \tilde{\mu}_{I\; \dot{2}}^{\; \dot{2}}) & -i \bar{\alpha} \bar d \psi_{\dot{1}-} -i \Lambda \bar{\beta}_{\perp} \psi_{\dot{2}+}\\
i(\alpha d +\Lambda \beta^{\perp}) \bar{\psi}_+^{\dot{1}} & A_{\varphi,I+1} +M_a{}^b \bar{r}_b r^a -\frac{i}{k} (\tilde{\mu}_{I+1\dot{1}}^{\quad\;\; \dot{1}} -\tilde{\mu}_{I+1\dot{2}}^{\quad\;\; \dot{2}})+c -\frac{1}{2}
\end{pmatrix},
\end{equation}
with $c=i \partial_{\varphi}\log \bar d$ and
\begin{equation}
M_a{}^{b}=\begin{pmatrix}
0 & \frac{i}{k \Lambda}\\
\beta^{\perp} \bar d \bar{\alpha} +(2 \bar d d -1) \frac{i}{k\Lambda} & \beta^{\parallel}
\end{pmatrix},
\end{equation}
where $M_{\parallel}{}^{\parallel}$ has been absorbed into $\beta^{\parallel}$, since both of them are free parameters. One can further absorb $\bar d$ into $\bar\alpha$ and $\bar\beta_\perp$, which sets $c=0$ and replaces $\alpha\to\alpha\bar d$ and $\beta^\perp\to\beta^\perp\bar d$. Then, with 
$\hat\beta^\perp=(\alpha (d\bar d-1)+\Lambda\beta^\perp\bar d)/\Lambda$ the bottom left entry in 
$\cL$ becomes $i (\alpha +\Lambda \hat\beta^{\perp})\bar{\psi}_+^{\dot{1}}$ and the 
bottom left entry in $M_a{}^{b}$ becomes $\bar\alpha\hat\beta^\perp+i/k\Lambda$.

This eliminates the parameters $d$ and $\bar d$ from $\cL$, so they are completely redundant. 
Furthermore, we see that these loops are exactly those found directly in the $\xi=0$ case in \eqref{xi=0} in 
Section~\ref{sec:2-node-xi}.

%%%%%%%%%%%%%

\bibliographystyle{utphys2}
\bibliography{refs}
\end{document}